\documentclass[%
 reprint,
superscriptaddress,
 amsmath,amssymb,
 aps,
]{revtex4-2}

\usepackage{graphicx}
\usepackage{dcolumn}
\usepackage{bm}
\usepackage{xcolor}

\begin{document}

\preprint{APS/123-QED}

\title{Self-coupling: An Effective Method to Mitigate Thermoacoustic Instability} 

\author{Sneha Srikanth}
\affiliation{%
Department of Mechanical Engineering, Indian Institute of Technology Madras, Chennai 600036, India}%
\author{Ankit Sahay}%
\author{Samadhan A. Pawar}
 \email{samadhanpawar@ymail.com}
 \affiliation{%
Department of Aerospace Engineering, Indian Institute of Technology Madras, Chennai 600036, India}%
\author{Krishna Manoj}
\affiliation{
Department of Mechanical Engineering, Massachusetts Institute of Technology, Cambridge, MA 02139, USA}
\author{R. I. Sujith}
\affiliation{%
Department of Aerospace Engineering, Indian Institute of Technology Madras, Chennai 600036, India}%

\date{\today}
\begin{abstract}
   The presence of undesirable large-amplitude self-sustained oscillations in combustors resulting from thermoacoustic instability can lead to performance loss and structural damage to components of gas turbine and rocket engines. Traditional feedback controls to mitigate thermoacoustic instability possess electromechanical components, which are expensive to maintain regularly and unreliable in the harsh environments of combustors. In this study, we demonstrate the quenching of thermoacoustic instability through self-coupling -- a method wherein a hollow tube is used to provide acoustic self-feedback to a thermoacoustic system. Through experiments and modeling, we identify the optimal coupling conditions for attaining amplitude death, i.e., complete suppression of thermoacoustic instabilities, in a horizontal Rijke tube. We examine the effect of both system and coupling parameters on the occurrence of amplitude death. We thereby show that the parametric regions of amplitude death occur when the coupling tube length is close to an odd multiple of the length of the Rijke tube. The optimal location of the coupling tube for achieving amplitude death is near the anti-node of the acoustic standing wave in the Rijke tube. Furthermore, we find that self-coupling mitigates thermoacoustic instability in a Rijke tube more effectively than mutual coupling of two identical Rijke tubes. Thus, we believe that self-coupling can prove to be a simple, cost-effective solution for mitigating thermoacoustic instability in gas turbine combustors.
\end{abstract}

\maketitle


\section{\label{sec:I}Introduction\protect}
The presence of undesirable self-sustained oscillations in many practical systems has been a matter of concern for researchers over the years. Notorious examples of such oscillations include wobbling bridges \cite{strogatz2005crowd}, fluttering aircraft wings \cite{garrick1981historical}, rumbling combustors \cite{lieuwen2005combustion,culick2006unsteady}, oscillatory prey-predator populations \cite{may2019stability,zou2021quenching}, and stock market fluctuations \cite{frankel2008adaptive}. The presence of these oscillations can have catastrophic consequences such as destruction of bridges, structural damage to aircraft and combustors, extinction of species, and financial crisis, respectively. It is thus vital to develop methods to quench these oscillations.

In the present study, we investigate the mitigation of thermoacoustic instability, which refers to the occurrence of ruinously high amplitude self-sustained oscillations in a combustor due to the reinforcing interaction between the heat release rate fluctuations from the heat source and the acoustic field of the combustor \cite{lieuwen2005combustion,sujith2020complex}. The manifestation of thermoacoustic instability can have several catastrophic consequences. It can cause thrust oscillations in rocket engines, which could jeopardize space missions. Thermoacoustic instability can cause rocket and gas turbine engine components to vibrate at high amplitudes, leading to catastrophic structural damage. The increased heat transfer due to thermoacoustic instability can overwhelm the thermal protection system and the large amplitude acoustic oscillations can damage electronic components in the engines \cite{poinsot2017prediction,sujith2020complex}.

Traditionally, control strategies to suppress thermoacoustic instability are classified into passive and active controls \cite{lieuwen2005combustion, mcmanus1993review,dowling2005feedback,poinsot2017prediction,sujith2021book}. Passive control strategies reduce the sensitivity of the combustor to acoustic disturbances by altering the geometry of the combustor or by installing components such as baffles and Helmholtz resonators \cite{lieuwen2005combustion,putnam1971combustion}. However, these strategies are effective only over a limited range of frequencies and the design changes involved in their implementation are costly and time-consuming. On the other hand, in most active control strategies, an actuator perturbs the system dynamics and breaks the coupling between the pressure and heat release rate fluctuations in the combustor, thereby leading to the suppression of thermoacoustic instability \cite{dowling2005feedback}. In this method, if the parameters of the forcing are controlled by the operator and are not directly dependent on the performance of the combustor, the control strategy is referred to as an open-loop control \cite{cosic2012open}; otherwise it is referred to as closed-loop control. 

In closed-loop control, also known as active feedback control, the pressure signal acquired from a thermoacoustic system is processed and used as input for a controller, which then accordingly instructs an actuator to modify system parameters (such as the inlet boundary conditions or the fuel flow rate). This, in turn, disrupts the flame-acoustic coupling and quenches thermoacoustic instability in the system \cite{heckl1988active, neumeier1996active}. The processing of the input pressure signal commonly involves phase-shifting and amplification. Though several studies have demonstrated the effectiveness of active feedback control in quenching thermoacoustic instability in various thermoacoustic systems \cite{heckl1988active,neumeier1996active,annaswamy1995active,dowling2005feedback}, the design, implementation, and maintenance of active feedback control systems is highly cumbersome. Moreover, the applicability of the feedback control method is restricted due to the lack of reliability of sensors and actuators for operating in the harsh environment of practical combustors, limited space for installing actuators, and high power requirements for operating feedback control units \cite{sujith2021book}. 

The aforementioned disadvantages of traditional passive and active control methodologies make it clear that there is a need to develop a control strategy that is cheap, simple to implement on practical combustion systems, and is effective in mitigating thermoacoustic instability. Towards this purpose, we propose the method of self-coupling in which we provide acoustic self-feedback using a tube. We refer to the quenching of limit cycle oscillations to a common fixed point due to the coupling of one or more oscillators as `amplitude death’ (AD) \cite{mirollo1990amplitude,reddy2000dynamics}. In self-coupling, the acoustic wave takes a finite time to propagate through the coupling tube and affect the acoustic field of the Rijke tube \cite{biwa2016suppression}. This leads to a delay in the self-feedback of acoustic oscillations of the system. Such delayed self-feedback has generally been implemented in the past to stabilize steady states in several systems, including the Van der Pol oscillator \cite{suchorsky2010using}, the Rössler oscillator \cite{ahlborn2005controlling}, and in electrochemical and optomechanical \cite{parmananda1999stabilization,naumann2014steady} systems. However, the practical implementation of these self-feedback techniques involve electronic components to capture the delayed signal and to amplify and feed it back to the system. This is resolved in the method of self-coupling used in our study where the acoustic pressure signal from a thermoacoustic system is directly fed to itself, without any explicit signal processing or modifying the inlet flow conditions. Thus, this method completely removes the requirement for any electro-mechanical components such as signal processors, sensors, and actuators used in traditional active and self-feedback controls. Additionally, the connecting tube is easier to design and implement as compared to traditional passive controls. Though methods similar to self-coupling, such as the Herschel-Quincke tube and the Infinity tube have been used in the past to suppress electrically driven acoustic pressure oscillations \cite{biwa2016suppression, lato2019passive}, their application in thermoacoustic systems is yet to be examined. 


In the present study, we address the following questions: (i) Can self-coupling mitigate thermoacoustic instability? (ii) What are the optimal values of coupling parameters (length, diameter, and location of the coupling tube) for attaining amplitude death in a thermoacoustic system? (iii) How does the amplitude of the limit cycle oscillations prior to introducing self-coupling affect its quenching? (iv) What is the nature of the transition between steady state and oscillatory state in the self-coupled thermoacoustic system? Towards this purpose, we systematically perform experimental and theoretical investigation on a horizontal Rijke tube subjected to self-coupling. 

The Rijke tube is a convenient prototype of a thermoacoustic oscillator traditionally used to study the occurrence and mitigation of thermoacoustic instability \cite{rijke1859lxxi,raun1993review,matveev2003thermoacoustic}. It consists of a duct with a heat source (e.g., an electrically heated wire mesh) present inside. The interaction between the heat released by the heat source and the acoustic field of the Rijke tube can lead to thermoacoustic instability. Through experiments, numerical simulations, and approximate analytical solutions, we demonstrate that self-coupling can mitigate thermoacoustic instability in a horizontal Rijke tube at optimal conditions. We observe that self-coupling causes amplitude death for wider parametric regions when the length of the coupling tube is close to an odd multiple of the length of the Rijke tube. The occurrence of amplitude death is easier for larger diameters of coupling tube and smaller amplitude of limit cycle oscillations prior to coupling. We show that the transition between the states of limit cycle oscillations and amplitude death due to self-coupling is explosive and hysteretic for a Rijke tube that exhibits subcritical Hopf bifurcation in the absence of coupling. We formulate a model for the self-coupled Rijke tube which qualitatively captures the experimental findings. We also find that the optimal coupling location for attaining amplitude death is around the centre of the Rijke tube, which corresponds to the antinode of the standing acoustic pressure wave developed in the Rijke tube prior to coupling. 

Recent studies have demonstrated the occurrence of amplitude death by mutual coupling of two Rijke tubes using one or two connecting tubes \cite{dange2019oscillation,sahay2021dynamics,hyodo2020suppression, srikanth2021dynamical}. Here, we also compare the effectiveness of suppressing limit cycle oscillations in a single Rijke tube via self-coupling against that obtained via mutual coupling of two identical Rijke tubes. We demonstrate that limit cycle oscillations of significantly greater amplitudes can be easily suppressed through self-coupling of a single Rijke tube as compared to mutual coupling of two such identical Rijke tubes. 

The rest of the paper is organized as follows. In Sec.~\ref{sec:II}, we provide details of the experimental setup. Following this, Sec.~\ref{sec:III} presents our results and discussions on the self-coupled Rijke tube. Within Sec.~\ref{sec:III} we experimentally investigate the effect of self-coupling on the mitigation of thermoacoustic instability in a horizontal Rijke tube in Sec.~\ref{sec:IIIA}. Subsequently, in Sec.~\ref{sec:IIIB}, we present a model of the self-coupled Rijke tube and investigate it numerically and analytically. In Sec.~\ref{sec:IIIC}, we compare the quenching of limit cycle oscillations due to the self-coupling of a Rijke tube with that obtained through mutual coupling of two such identical Rijke tubes. Finally, we present our conclusions from the study in Sec.~\ref{sec:IV}. 

\section{\label{sec:II} Experimental setup of the self-coupled Rijke tube} 

\begin{figure}
\centering
\includegraphics[width=8.6cm]{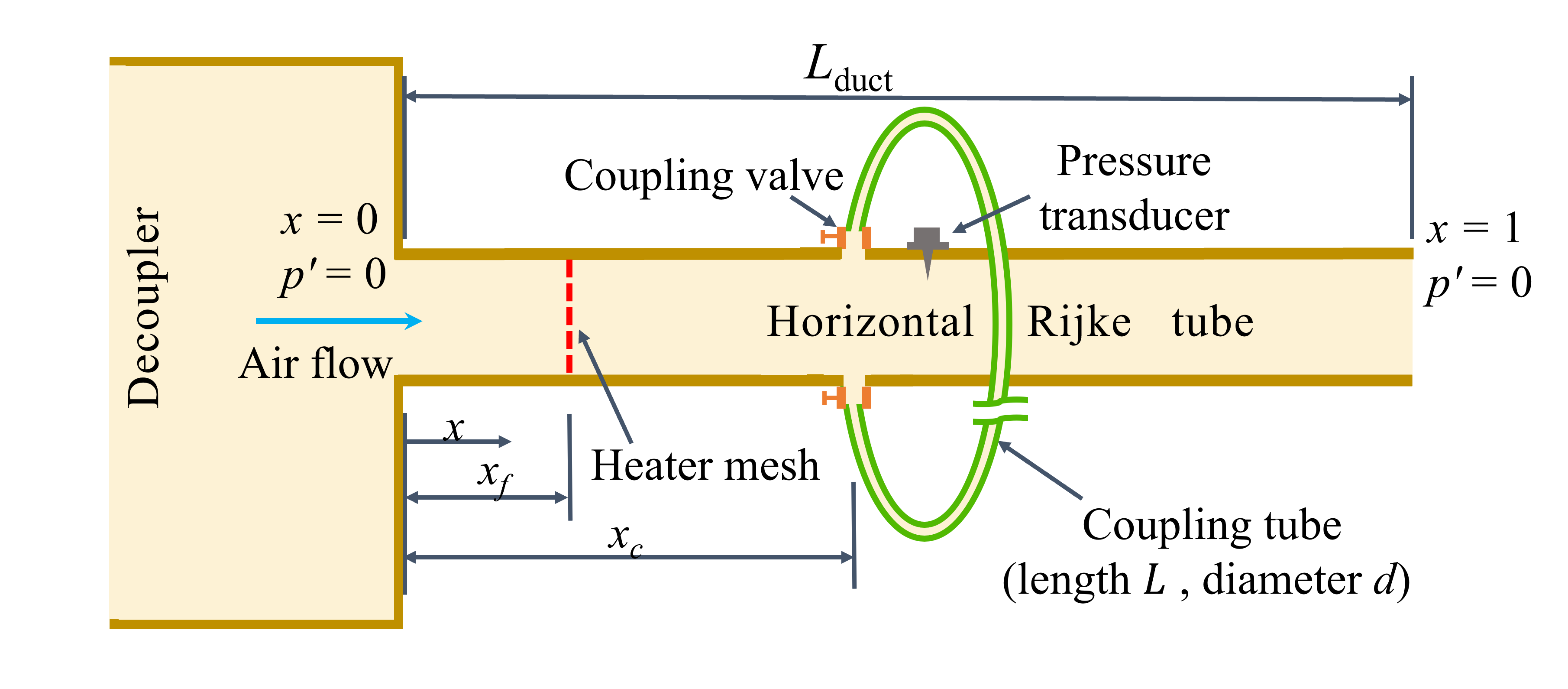}
\caption{\label{setup} Schematic of a horizontal Rijke tube self-coupled using a connecting tube. The axial location (normalized by the length of the Rijke tube, $L_\text{duct}$) of the heater mesh and the connecting tube are denoted by $x_f$ and $x_c$, respectively. Ball-type coupling valves are manually opened to establish acoustic feedback in the system.
}
\end{figure}

In Fig.~\ref{setup}, we show a schematic representation of the horizontal Rijke tube used in the present study. The Rijke tube is a long duct with a rectangular cross section of $ 9.3$ cm $\times$ 9.4 cm and a length ($L_\text{duct}$) of $104$ cm, similar to the ones used in \cite{gopalakrishnan2014influence,dange2019oscillation,sahay2021dynamics}. An electrically heated wire mesh, powered by an external DC power supply, acts as a compact heat source. The heated wire mesh is located at $27$ cm downstream from the inlet in the duct. Air flow is supplied to the Rijke tube through a mass flow controller (MFC, Alicat Scientific) of uncertainty $\pm$(0.8\% of the measured reading + 0.2\% of the full scale reading). A decoupler of dimensions 102 cm $\times$ 45 cm $\times$ 45 cm is attached to the inlet of the Rijke tube. The decoupler is a large chamber used to dampen out the fluctuations in the incoming air flow so that a steady air flow enters the Rijke tube. During experiments, the heater power supplied to the wire mesh is increased so that the system behavior transitions from steady state to limit cycle oscillations via subcritical Hopf bifurcation for the given air flow rate \cite{etikyala2017change}. 

In order to quench the limit cycle oscillations, we implement self-coupling to acoustically couple the Rijke tube to itself using a single vinyl tube of length $L$ and internal diameter $d$. The self-coupling tube is attached to two sides of the Rijke tube at an axial distance of $57$ cm from the inlet of the Rijke tube (the axial location of the connecting tube on normalizing by $L_\text{duct}$ is $x_c = 0.55$), unless otherwise specified. Ball-type valves are manually operated to control the self-coupling in the system. The length of the connecting tube ($L$) is varied from 92 cm to 362 cm in steps of 5 cm. Coupling tubes of diameters ($d$) 4 mm to 12 mm in steps of 2 mm are considered in this study. 

To measure the acoustic pressure fluctuations in the system, a piezoelectric pressure transducer (PCB 103B02, sensitivity $217.5$ mV/$\text{kPa}$, uncertainty $\pm 0.15$ Pa) is mounted along the length of the duct at 57 cm from the inlet. The pressure data is acquired from the Rijke tube at a sampling rate of $10$ kHz for a duration of $3$ s for each parametric condition using a data acquisition system (NI USB 6343). The resolution of frequency in the power spectrum of the signal is equal to 0.2 Hz. To measure the acoustic damping in the Rijke tube, we send an acoustic pulse into the Rijke tube using a loudspeaker in the absence of air flow and heating and calculate its decay rate. We observe the decay rate for the Rijke tube to be $15.8 \pm 2$ $\text{s}^{-1}$. To ensure consistency of the experimental conditions and repeatability of the results, the experiments are conducted only when the measured acoustic decay rates lie in the aforementioned range. 

Before starting any experiment, the Rijke tube is preheated for $10$ minutes in the steady state regime of operation by supplying DC power at $1$ V to the wire mesh. The preheating ensures a steady temperature profile inside the Rijke tube. The experimental setup for the mutually coupled Rijke tubes (discussed in Sec.~\ref{sec:IIIC}) is similar to the above description, except that we couple two identical Rijke tubes with the connecting tube. For more experimental details on mutually coupled Rijke tubes, readers may refer to \cite{dange2019oscillation,sahay2021dynamics}.

\section{\label{sec:III}Results and Discussion}

\subsection{\label{sec:IIIA} Experimental investigation of the self-coupled Rijke tube} 

\begin{figure}[b]
\centering
\includegraphics[width=8.6cm]{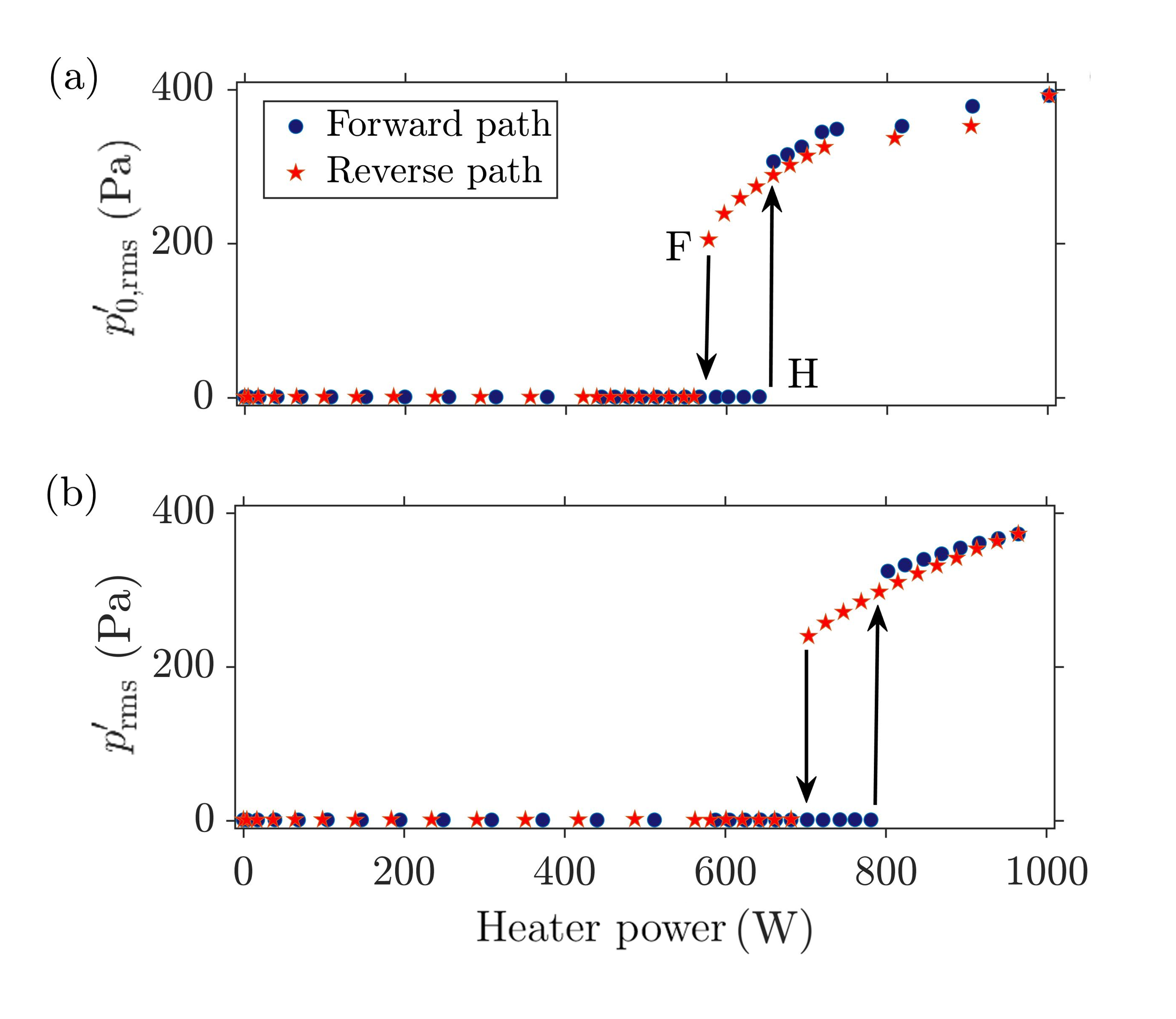}
\caption{\label{fig4} Variation of the RMS value of acoustic pressure fluctuations with heater power in the forward and reverse paths for (a) the uncoupled Rijke tube ($p^\prime_\text{0,rms}$) and (b) the self-coupled Rijke tube ($p^\prime_\text{rms}$). The normalized length of the connecting tube is $L/L_\text{duct}=1.17$ and its internal diameter is $d=8$ mm in (b). The introduction of self-coupling shifts the Hopf point [marked as `H' in (a)] and the fold point [marked as `F' in (a)] to higher values of heater power. Air flow rate of 120 SLPM (or 0.002 m\textsuperscript{3}/s) is maintained in the Rijke tube for both the plots.}
\end{figure}

\begin{figure*}
\centering
\includegraphics[width=14cm]{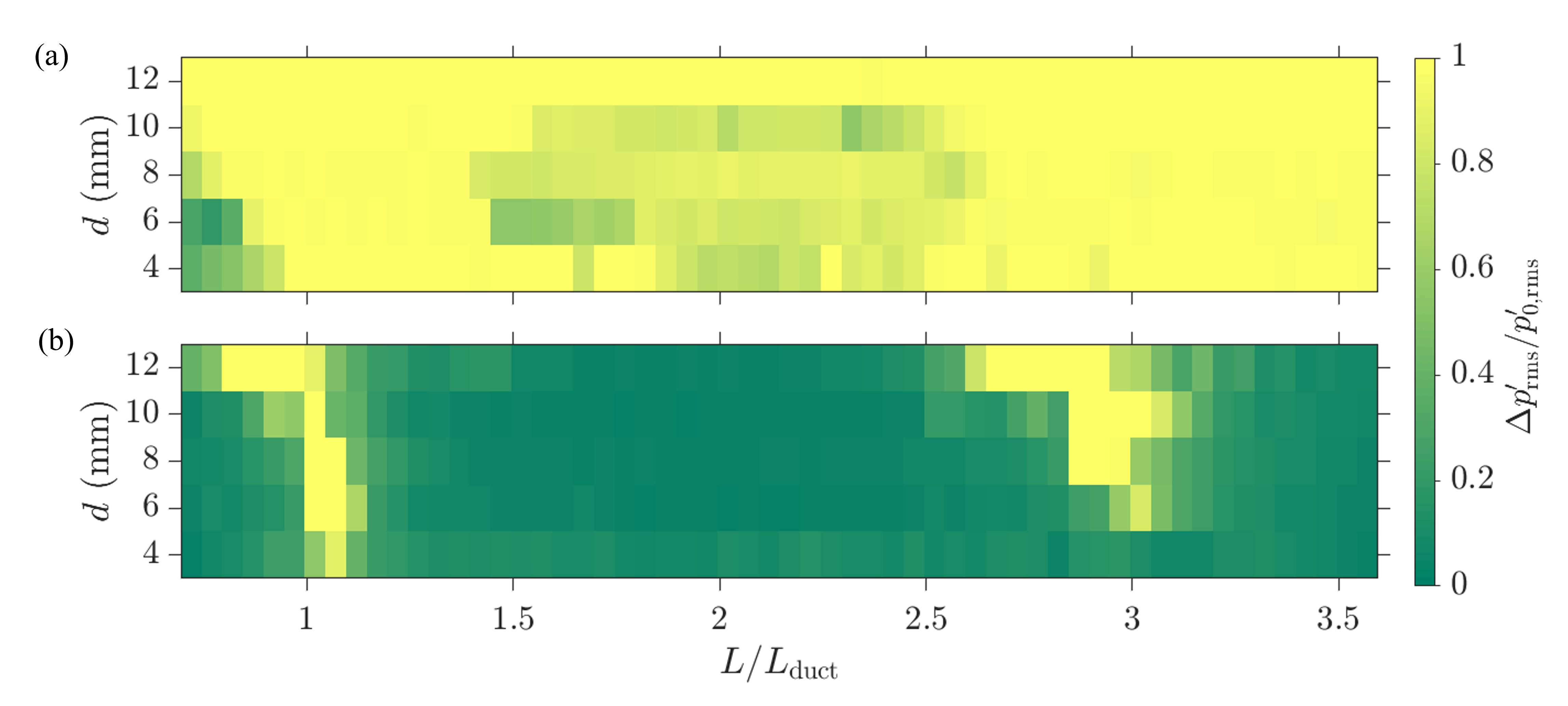}
\caption{\label{fig2} The amplitude response of the Rijke tube when self-coupling is induced for (a) $p^\prime_\text{0,rms}=120$ Pa, and (b) $p^\prime_\text{0,rms}=320$ Pa. The normalized change in the amplitude suppression $\Delta p^\prime_\text{rms}/p^\prime_\text{0,rms}$ is measured for different values of $L/L_\text{duct}$ and $d$. A large parametric region of amplitude death (AD) is observed for lower value of $p'_\text{0,rms}$ in (a); while for higher values of $p^\prime_\text{0,rms}$ in (b), the region of amplitude death shrink and is observed only for around $L/L_\text{duct} \approx 1$ and $L/L_\text{duct} \approx 3$. Air flow rate is maintained at 80 SLPM (or 0.00133 m\textsuperscript{3}/s) in the Rijke tube.}
\end{figure*}

In this section, we first discuss the effect of self-coupling on the bifurcation characteristics of the Rijke tube as observed in our experiments. In Fig. \ref{fig4}(a), we show the variation in the root-mean-square (RMS) value of acoustic pressure fluctuations (in Pa) as a function of the heater power (in W) in the absence of self-coupling in the system. We denote the RMS value using $p^\prime_\text{0,rms}$ for the uncoupled case and $p^\prime_\text{rms}$ for the coupled case. In the forward path (increasing heater power), we observe that the transition from fixed point to limit cycle oscillations of RMS value around 290 Pa happens via subcritical Hopf bifurcation; while in the reverse path (decreasing heater power), the system attains fixed point via fold bifurcation when $p^\prime_\text{0,rms}$ is around 200 Pa. Thus, the transition between the steady state and the limit cycle oscillatory state is explosive (subcritical) and hysteretic. In Fig. \ref{fig4}(a), `H' denotes the Hopf bifurcation point and `F' denotes the fold bifurcation point. We notice a bistable region in between the two bifurcation points. When the Rijke tube is self-coupled with a single coupling tube of dimensions $L/L_\text{duct}=1.17$ and $d=8$ mm in Fig. \ref{fig4}(b), we observe that the system preserves the bifurcation characteristics that we observed in the uncoupled case [see Fig. \ref{fig4}(a)]. However, we notice that the introduction of self-coupling shifts the Hopf and the fold bifurcation points to higher values of the heater power than those values observed for the uncoupled oscillator. This, in turn, indicates that self-coupling enhances the parameter space of steady state in the system; in other words, it delays the transition to limit cycle oscillations in the system. 

\begin{figure*}[t]
\centering
\includegraphics[width=15cm]{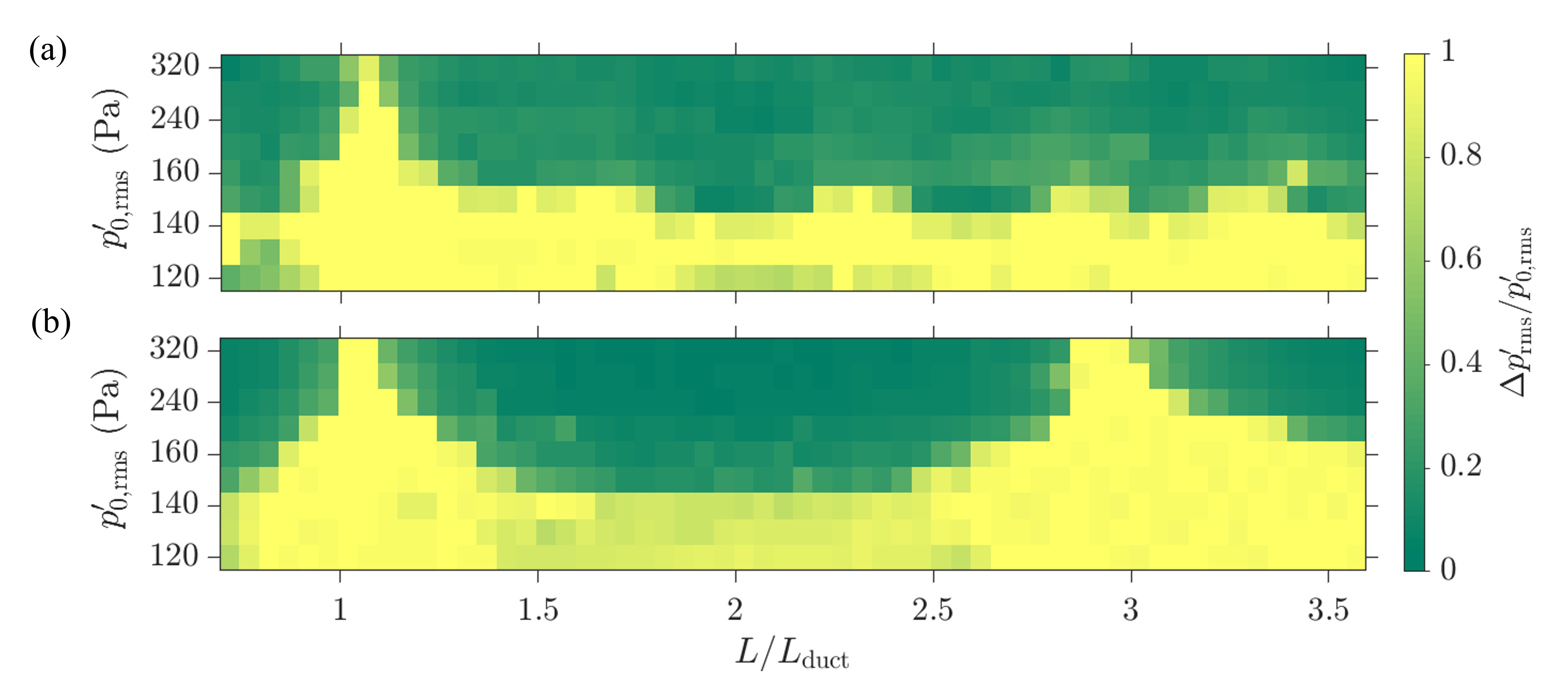}
\caption{\label{fig:p0_L} The suppression of acoustic pressure fluctuations in the self-coupled Rijke tube is shown for different values of the normalized length of connecting tube ($L/L_\text{duct}$) and the RMS value of oscillations in the uncoupled state ($p'_\text{0,rms}$) when the internal diameter of coupling tube ($d$) is fixed at (a) $4$ mm, and (b) $8$ mm. Air flow rate is maintained at 80 SLPM (or 0.00133 m\textsuperscript{3}/s) in the Rijke tube.}
\end{figure*}

Next, we study the effect of variation in the parameters of the connecting tube (i.e., length $L$ and diameter $d$) and the RMS value of limit cycle oscillations in the uncoupled state ($p'_\text{0,rms}$) on the suppression characteristics in the self-coupled Rijke tube oscillator. We note that prior to implementing self-coupling, we establish limit cycle oscillations in the system. As discussed in Sec.~\ref{sec:II}, the value of $p'_\text{0,rms}$ can be increased by increasing the value of the heater power. To study the behavior of the self-coupled Rijke tube over a larger range of $p'_\text{0,rms}$ than that shown in Fig.~\ref{fig4}(a), we maintain the air flow rate at a value of 80 SLPM (or 0.00133 m\textsuperscript{3}/s) in our subsequent experiments. Lowering the air flow rate reduces the value of $p'_\text{0,rms}$ at the Hopf point of the Rijke tube \cite{etikyala2017change}. At this air flow rate, we vary the value of $p'_\text{0,rms}$ for a larger range of 120 Pa to 320 Pa by varying the heater power. We refer to the dimensions of the coupling tube (i.e., $L$ and $d$) as coupling parameters and the RMS value ($p'_\text{0,rms}$) and frequency ($f_0$) of limit cycle oscillations prior to coupling as system parameters in the subsequent discussion of the paper.  


Figure~\ref{fig2} illustrates the amplitude response of the self-coupled Rijke tube for different coupling parameters. Here, we show the amplitude response in terms of the normalized change in the RMS value of limit cycle oscillations in the Rijke tube due to self-coupling, $\Delta p^\prime_\text{rms}/p^\prime_\text{0,rms} = (p^\prime_\text{0,rms} - p^\prime_\text{rms})/p^\prime_\text{0,rms}$, as a function of $L/L_\text{duct}$ and $d$. The color bar illustrates values of $\Delta p^\prime_\text{rms}/p^\prime_\text{0,rms}$ ranging from $0$ to $1$, where $\Delta p^\prime_\text{rms}/p'_\text{0,rms} \approx 1$ corresponds to complete suppression of limit cycle oscillations (i.e., amplitude death) and $\Delta p^\prime_\text{rms}/p'_\text{0,rms} \approx 0$ indicates the absence of any suppression of limit cycle oscillations in the Rijke tube due to self-coupling. We study the response of the self-coupled Rijke tube oscillator for two different RMS values of limit cycle oscillations in the uncoupled state: $p'_\text{0,rms} = 120$ Pa [see Fig.~\ref{fig2}(a)], and $p'_\text{0,rms} = 320$ Pa [see Fig. \ref{fig2}(b)].

In Fig.~\ref{fig2}(a), we note that when the RMS value of the limit cycle oscillations is low ($p'_\text{0,rms} = 120$ Pa), self-coupling of the Rijke tube causes a significant reduction in the amplitude of acoustic pressure oscillations for a vast parametric region of coupling parameters. However, as the value of $p'_\text{0,rms}$ is increased to a higher value of $320$ Pa in Fig.~\ref{fig2}(b), we observe the region of amplitude death only around odd values $L/L_\text{duct}$ (i.e., $L/L_\text{duct} \approx 1$ and $3$). We also notice that the range of $L/L_\text{duct}$ for which the Rijke tube exhibits amplitude death increases as the internal diameter of the coupling tube ($d$) is increased.

\begin{figure*}[t]
\centering
\includegraphics[width=12cm]{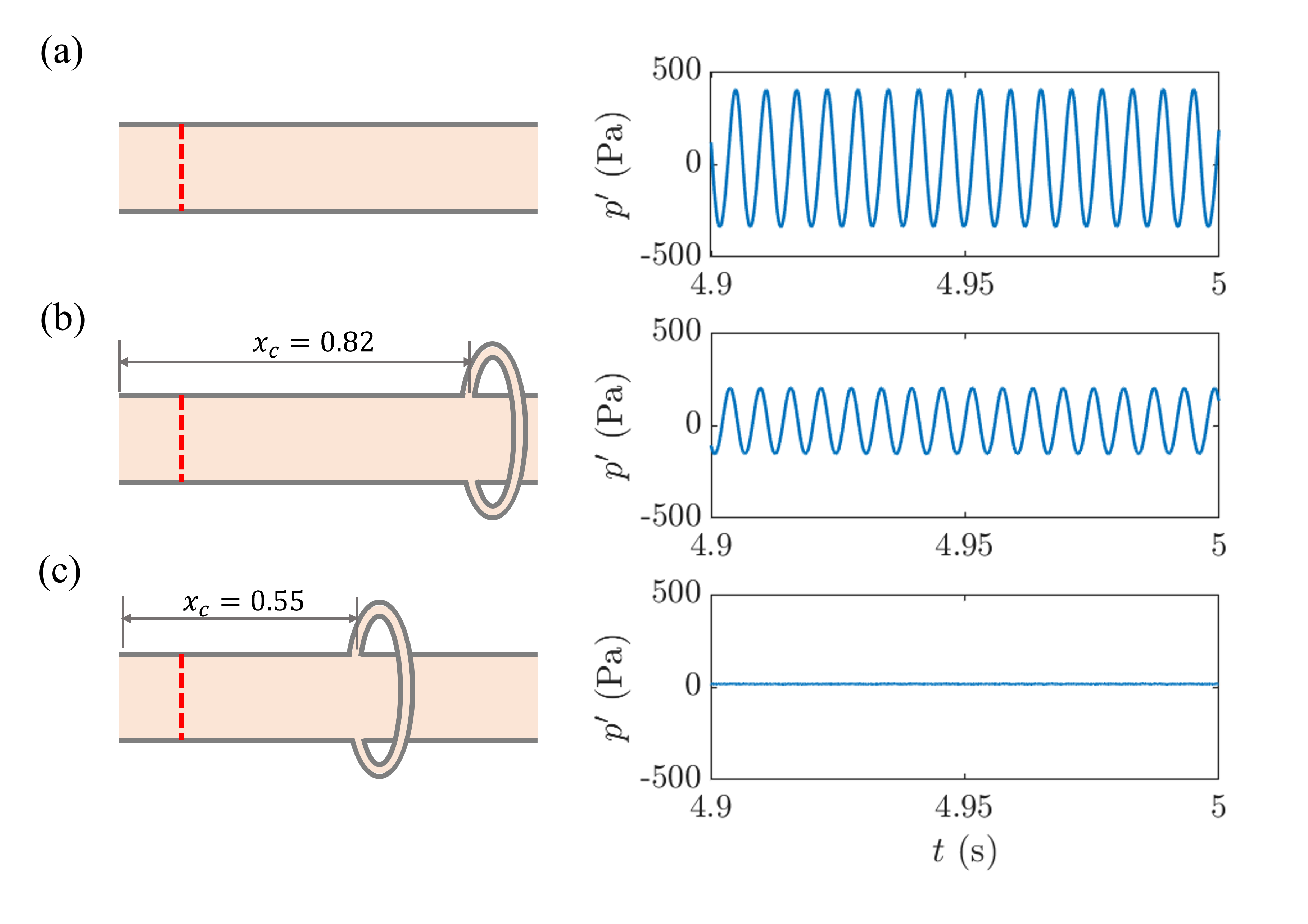}
\caption{\label{fig:expt_loc} Schematics of the Rijke tube and the time series of the acoustic pressure fluctuations ($p'$) depict the effect of coupling location on the suppression of thermoacoustic instability for: (a) the uncoupled Rijke tube, (b,c) and the self-coupled Rijke tube when the coupling tube is fixed at a distance ($x_c$) of 0.82 and 0.55 from the inlet, respectively. All distances are non-dimensionalized by the length of the Rijke tube, $L_\text{duct} = 1.04$ m. Other parameters are maintained as $p'_\text{0,rms} = 200$ Pa, $L_\text{duct} = 95$ cm, and $d = 4$ mm in all of the plots.}
\end{figure*}

Next, we examine in detail the amplitude response of the self-coupled Rijke tube on variation of the RMS value of the limit cycle oscillations in the uncoupled state ($p'_\text{0,rms}$) and the normalized length of the connecting tube ($L/L_\text{duct}$). The internal diameter of the coupling tube is kept constant at $4$ mm and $8$ mm in Fig. \ref{fig:p0_L}(a) and \ref{fig:p0_L}(b), respectively. Note that the ordinate representing the $p'_\text{0,rms}$ values is unevenly distributed. In Fig. \ref{fig:p0_L}, we observe that limit cycle oscillations of low RMS values (i.e., for $p'_\text{0,rms} \leq 140$ Pa), and thus low amplitudes, can be easily suppressed for a large range of $L/L_\text{duct}$. As $p'_\text{0,rms}$ is increased above $140$ Pa, we notice that the amplitude death region shrinks and occurs only in a narrow range of $L/L_\text{duct}$. For a low value of $d$ in Fig. \ref{fig:p0_L}(a), we observe amplitude death to occur near $L/L_\text{duct}=1$ when $p'_\text{0,rms}$ is greater than 140 Pa. In contrast, we notice two distinct regions of amplitude death in Fig. \ref{fig:p0_L}(b) when the value of $d$ is increased to $8$ mm. Here, the two regions of amplitude death occur at around $L/L_\text{duct} = 1$ and $3$, respectively, which is similar to what we observed in Fig. \ref{fig2}(b). Thus, from Figs.~\ref{fig2} and \ref{fig:p0_L}, we infer that self-coupling can quench small amplitude limit cycle oscillations for a large range of coupling parameters. On the other hand, the quenching of large amplitude limit cycle oscillations is possible only for a critical range of the length of the coupling tube and it is comparatively easy when the diameter of the coupling tube is bigger. 

So far, we have investigated the behavior of the self-coupled Rijke tube for a fixed location of the coupling tube. We next examine how varying the coupling location affects the quenching of limit cycle oscillations by tracking the temporal variation of acoustic pressure oscillations in the system. We observe that as compared to the uncoupled state [depicted in Fig.~\ref{fig:expt_loc}(a)], coupling the Rijke tube at a location towards the end of the duct does not cause significant suppression of limit cycle oscillations [Fig.~\ref{fig:expt_loc}(b)]. On the other hand, when the location of the coupling tube along the duct of the Rijke tube is closer to the center [Fig.~\ref{fig:expt_loc}(c)], the oscillations are quenched and amplitude death is achieved. This suggests that the optimal axial location of the coupling tube for mitigating thermoacoustic instability is around the center of the Rijke tube. We discuss this behavior of the self-coupled Rijke tube through modeling in the next section.
\subsection{\label{sec:IIIB} Theoretical analysis of the self-coupled Rijke tube oscillator} 

In this section, we theoretically analyze a model of the self-coupled Rijke tube oscillator. We build on the reduced-order model of the horizontal Rijke tube proposed by Balasubramanian and Sujith \cite{balasubramanian2008thermoacoustic}. We derive the governing equations of the self-coupled Rijke tube in the following manner. First, we consider the non-dimensionalized linearized momentum and energy equations for the acoustic field of the duct by neglecting the effect of mean flow (zero Mach number approximation \cite{nicoud2009zero}) and mean temperature gradient:
\begin{align}
    \gamma M \dfrac{\partial u'}{\partial t} + \dfrac{\partial p'}{\partial x} &= 0,\label{eq1}\\
    \dfrac{\partial p'}{\partial t} +  \gamma M\dfrac{\partial u'}{\partial x} + \zeta p' &= (\gamma-1)\dot{Q}'\delta(x-x_f) + C\delta(x-x_c).
    \label{eq2}
\end{align}
Here, $x$ is the axial distance non-dimensionalized by the length of the duct $L_\text{duct}$. $t$ is the time non-dimensionalized by the ratio $L_\text{duct}/c_0$, where $c_0$ is the speed of sound. $u'$ and $p'$ are the acoustic velocity and acoustic pressure, non-dimensionalized by the steady state velocity ($u_0$) and pressure ($p_0$), respectively, of the air flowing through the Rijke tube. $\gamma$ is the ratio of specific heats, while $M$ is the Mach number ($M=u_0/c_0$). $\zeta$ captures the damping in the model. $\dot{Q}'$ is the non-dimensional heat release rate fluctuations per unit area from the heat source. $C$ is the self-coupling term added to account for the effects induced by the connecting tube. $\dot{Q}'$ is multiplied by a Dirac delta function to indicate that the heat source is concentrated at the non-dimensional heater location $x_f$ in the Rijke tube. Similarly, the two ends of the connecting tube are located at the same coupling location $x_c$ along the axial length of the Rijke tube. We formulate $\dot{Q}'$ using the correlation given by Heckl
\cite{heckl1990non}:
\begin{align}
    \dot{Q}' = \frac{2 L_w (T_w - T_0)}{c_0p_0 S \sqrt{3}}  &\sqrt{\pi \lambda_T C_v u_0 \rho_0 r_w}\nonumber\\
    &\times \left[ \sqrt{\left\lvert\frac{1}{3} + u'(x,t-\tau_h) \right\rvert} - \sqrt{\frac{1}{3}} \right], \label{eq3}
\end{align}
where $r_w$, $T_w$, and $L_w$  are the radius, the temperature, and the length of the heated wire, respectively. $T_0$ and $\rho_0$ are the steady state temperature and density of the medium. $S$ is the cross-sectional area of the duct. $\lambda_T$ and $C_v$ are the thermal conductivity and the specific heat at constant volume of the medium, respectively. $u'(x,t-\tau_h)$ is the acoustic velocity at time $t-\tau_h$. Here, the time lag, $\tau_h$, accounts for the thermal inertia of heat transfer in the medium \cite{lighthill1954response}.

We use time-delay coupling to capture the effect of delayed interaction of acoustic waves propagating through the connecting tube with the acoustic field of the duct. The coupling term is described by the following expression:
\begin{align}
    C = K_\tau \left[p'(x, t - \tau) - p'(x,t) \right], \label{eq4}
\end{align}
where $K_\tau$ is the coupling strength and $\tau$ is the coupling delay characterizing the time it takes for the acoustic waves to propagate through the connecting tube. 

The normalized length of the connecting tube ($L/L_\text{duct}$) in the experiment can be related to the coupling delay ($\tau$) in the model in the following manner:
\begin{align}
    \tau &= \dfrac{\tilde{\tau}}{L_\text{duct}/c_0},
\intertext{where $\tilde{\tau}$ is the dimensional coupling delay. Assuming that the acoustic waves propagate through the connecting tube at speed $c_0$, the time taken by the sound waves to travel through connecting tube is}
\tilde{\tau} &= L/c_0.
\intertext{Hence,}
\tau &= \dfrac{L/c_0}{L_\text{duct}/c_0} = L/L_\text{duct}. \label{eq:relation_tau_L}
\end{align}
Substituting Eq.~(\ref{eq3}) and (\ref{eq4}) into the energy equation Eq.~\ref{eq2}, we get the following equation:
\begin{align}
    \dfrac{\partial p'}{\partial t} &+  \gamma M\dfrac{\partial u'}{\partial x} + \zeta p' \nonumber\\&= \frac{K}{2}\left[ \sqrt{\left\lvert\frac{1}{3} + u'(t-\tau_h) \right\rvert} - \sqrt{\frac{1}{3}} \right]\delta(x-x_f) 
    \nonumber\\&\quad + K_\tau \left[p'(x, t - \tau) - p'(x,t) \right]\delta(x-x_c), \label{eq5}
\end{align}
where $K$ is the non-dimensional heater power given by:
\begin{align}
    K = \frac{4(\gamma - 1)L_w}{\gamma M c_0 p_0 S \sqrt{3}} (T_w - T_0) \sqrt{\pi \lambda_T C_v u_0 \rho_0 r_w}.
\end{align}

We now employ the Galerkin technique \cite{lores1973nonlinear} to decompose the partial differential equations given by Eq.~(\ref{eq1}) and (\ref{eq5}) into a set of ordinary differential equations. To that end, we choose the basis functions of $u'$ and $p'$ to be the natural acoustic modes of the duct in the absence of the heater \cite{balasubramanian2008thermoacoustic}. The duct of the Rijke tube is open at both ends, where the total pressure ($p$) must equal the ambient pressure ($p_0$). As a result, the acoustic pressure fluctuations ($p' = p-p_0$) are absent at the ends of the tubes. So, the basis functions must satisfy the boundary conditions $p'(0,t) = p'(1,t) = 0$. Thus, $u'$ and $p'$ are decomposed into their modes as:
\begin{align}
    u'(x,t) = \sum_{j=1}^N \eta_j \cos(j\pi x), \label{eq7}\\
    p'(x,t) = -\sum_{j=1}^N \frac{\gamma M}{j\pi} \dot{\eta}_j \sin(j\pi x), \label{eq8}
\end{align}
where $\eta_j$ and $\dot{\eta}_j$ are the time-varying coefficients of the $j$\textsuperscript{th} mode of $u'$ and $p'$, respectively, and $N$ is the total number of Galerkin modes considered in the expansion. On employing the Galerkin technique, we obtain the set of ordinary differential equations from the partial differential equations Eq.~(\ref{eq1}) and (\ref{eq5}) as:
\begin{align}
    \ddot{\eta}_j &+ 2\zeta j\pi \dot{\eta}_j + (j\pi)^2 \eta_j \nonumber\\
    = &- j\pi K\left[ \sqrt{\left\lvert\frac{1}{3} + u'_f(t-\tau_h) \right\rvert} - \sqrt{\frac{1}{3}} \right]\sin(j\pi x_f) \nonumber\\
    & - \frac{j\pi}{\gamma M} K_\tau \left[p'_c( t - \tau) - p'_c(t) \right]\sin(j \pi x_c), \label{eq9}
\end{align}
where $u'_f(t-\tau_h) = u'(x_f, t-\tau_h)$ and $p'_c(t-\tau) = p'(x_c, t-\tau)$. $\zeta_j$ is the frequency dependent damping given by \cite{matveev2003thermoacoustic}:
\begin{equation}
    \zeta_j = \dfrac{1}{2\pi} \left( c_1 \dfrac{\omega_j}{\omega_1} + c_2 \sqrt{\dfrac{\omega_1}{\omega_j}} \right),
\end{equation}
where $c_1$ and $c_2$ are the damping coefficients. Equation~(\ref{eq9}) represents the set of governing equations for the self-coupled Rijke tube oscillator. 

Before performing numerical simulations on the model, we attempt to derive the conditions for attaining amplitude death in the system analytically. Towards this purpose, we determine the stability of the steady state in the system and locate the parameter values where the solution $\eta_j = 0$ loses its stability. We first simplify Eq.~(\ref{eq9}) by considering only the first mode and linearizing the heat release rate term under the assumption of small amplitudes \cite{subramanian2013subcritical}. We subsequently drop the subscript $j$ and perform algebraic manipulations to yield the following differential equation:
\begin{align}
    \ddot{\eta} + 2\zeta \pi \dot{\eta} + \pi^2 \eta = - \frac{\sqrt{3}}{4} K\sin(2\pi x_f)\eta(t-\tau_h) \nonumber\\+ K_\tau \left[\dot{\eta}(t - \tau) - \dot{\eta}(t) \right]\sin^2(\pi x_c). \label{eq10}
\end{align}

\begin{figure*}[t]
\includegraphics[width= 13cm]{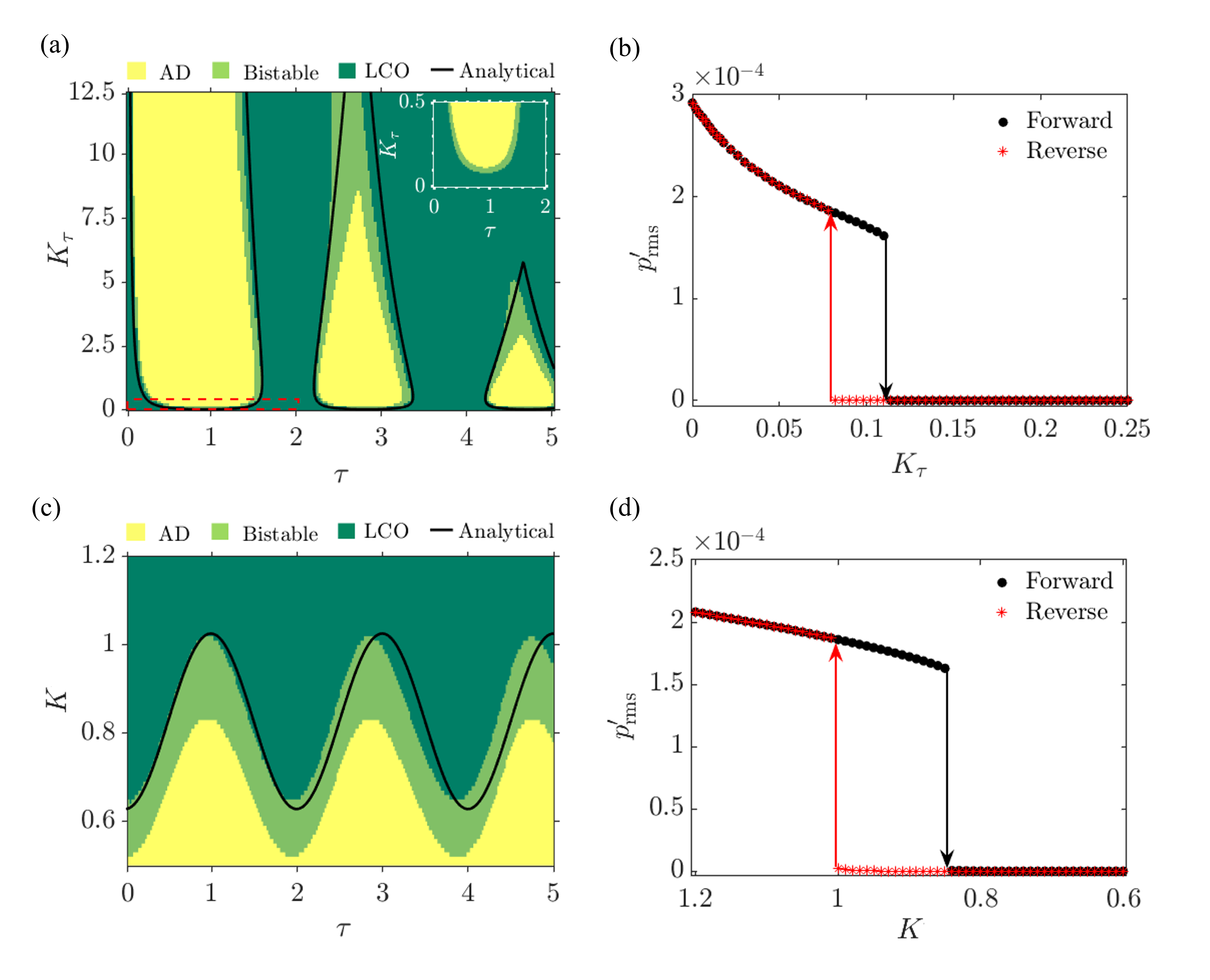}
\caption{\label{fig:twoparam}
(a) Two-parameter bifurcation plot between coupling strength ($K_\tau$) and coupling delay ($\tau$) showing multiple regions of amplitude death (AD) in the model of the self-coupled Rijke tube oscillator. The black curve denotes the boundary between the regions of AD and limit cycle oscillatory state (LCO) determined using linear stability analysis. Inset containing a part of the bifurcation plot shows the presence of LCO for low coupling strength. (b) One-parameter bifurcation plot between $p'_\text{rms}$ and $K_\tau$ for $\tau= 1$ illustrates the occurrence of explosive transition and hysteresis between LCO and AD in the system. The non-dimensional heater power $K$ is maintained at 1.2 in (a) and (b). (c) Two-parameter bifurcation between $K$ and $\tau$ shows the oscillating boundary of the amplitude death region with $\tau$ in the system. The black curve denotes the boundary between the AD and LCO regions obtained analytically using the method of averaging. (d) One-parameter bifurcation plot between $p'_\text{rms}$ and $K$ for $\tau= 1$. $K_\tau$ is maintained at 0.05 in (c) and (d).}
\end{figure*}

We now substitute a solution of the form $\eta = e^{qt}$ into Eq.~(\ref{eq10}), where $q = p + i\omega$ is an eigenvalue of the differential equation. Here, $p$ is the real part of the eigenvalue and it describes the rate at which perturbations to the steady state decay or grow. $\omega$ is the imaginary part of the eigenvalue and it describes the angular frequency with which the perturbations oscillate while decaying or growing in magnitude \cite{lakshmanan2011dynamics}. Thus, when $p<0$, the steady state is stable in the self-coupled system and so amplitude death can be achieved. In this manner, we detect the critical parameter values for which $p$ changes its sign (i.e., crossing $p=0$) and obtain the expressions for the boundary demarcating the amplitude death region as:  
\begin{align}
    \tan\left(\frac{\omega \tau}{2}\right) &= \dfrac{\sqrt{3} K\pi\sin(2\pi x_f)\sin(\omega \tau_h) - 8\pi\zeta \omega}
    {\sqrt{3} K\pi\sin(2\pi x_f)\cos(\omega \tau_h) + 4(\pi^2 - \omega^2)}, \label{eq11} \\
    K_\tau &= \frac{8\pi\zeta \omega - \sqrt{3}K\pi\sin(2\pi x_f)\sin(\omega \tau_h)}{4\sin^2(\pi x_c)\left(\omega \cos(\omega \tau) - 1 \right)}, 
    \label{eq12}
\end{align}
where $\omega$ is the imaginary part of $q$. Since the natural frequency of the first mode of the duct in the absence of the heater is $\pi$, we vary the value of $\omega$ around this value to yield the parametric curve describing the boundary of the amplitude death region, as illustrated later on in Fig.~\ref{fig:twoparam}(a).

To further simplify the conditions for obtaining AD, we employ the method of averaging on the linearized equation Eq.~(\ref{eq10}) by assuming additionally small values of $K$, $K_\tau$, $\tau_h$, and $\tau$. This gives us the condition for amplitude death as:
\begin{align}
    \cos(\omega \tau) < 1-\dfrac{\sqrt{3} K\sin(2\pi x_f) \sin(\omega \tau_h) - 8\pi \zeta }{4K_\tau \sin^2(\pi x_c)}.  \label{eq13}
\end{align}
We use the above condition subsequently in Fig.~\ref{fig:twoparam}(c) to demarcate the parametric regions of amplitude death and compare it with our numerical and experimental results. 

Next, we discuss the effect of self-coupling on the occurrence of amplitude death in the model of the horizontal Rijke tube. We primarily vary four parameters in the model: (i) coupling strength ($K_\tau$), (ii) coupling delay ($\tau$), (iii) non-dimensional heater power ($K$), and (iv) coupling location ($x_c$). We utilize the approximate analytical solutions shown in Eqs.~(\ref{eq11})-(\ref{eq13}) to visualize the optimal conditions for achieving amplitude death in the self-coupled Rijke tube. We also perform numerical simulations by numerically integrating the governing equation of the model Eq.~(\ref{eq9}) using the inbuilt function dde23 of MATLAB$^{\tiny{\text{\textregistered}}}$
 \cite{shampine2000solving}. In all of our simulations, we measure the dynamics of the acoustic pressure fluctuations in the Rijke tube at the heater location $x_f$. We use the first 5 modes for our numerical simulations since we observe negligible changes in the dynamics of the system on consideration of higher modes \cite{subramanian2010bifurcation}. In Eq.~(\ref{eq9}), we fix the values of $\tau_h$, $\gamma$, $c_0$, $c_1$ and $c_2$ at 0.2, 1.4, 340 m/s, 0.1 and 0.06, respectively, based on previous theoretical studies \cite{subramanian2010bifurcation, thomas2018effect}. According to the experiments in the present study, we maintain the value of $M$ at $4.49 \times 10^{-4}$ and $x_f$ at $0.256$. We restrict the range of our system parameter $K$ to 1.2 so that we primarily excite the first mode of the Rijke tube and obtain period-1 limit cycle oscillations in the uncoupled state of the system.

First, we examine the effect of varying the coupling strength ($K_\tau$) and coupling delay ($\tau$) on the dynamical behavior of the self-coupled Rijke tube. In Fig.~\ref{fig:twoparam}(a), we show the two-parameter bifurcation diagram between the coupling parameters $K_\tau$ and $\tau$ for a high value of non-dimensional heater power ($K=1.2$). The coupling location $x_c$ is fixed at $0.55$, which is the same as in our experiments. We observe that self-coupling can subdue high amplitude thermoacoustic oscillations in the Rijke tube for particular ranges of coupling delay and these ranges are roughly centered around odd numbers of $\tau$. Additionally, we notice that the islands of amplitude death decrease in size with an increase $\tau$. We compare Figs.~\ref{fig2}(b) with \ref{fig:twoparam}(a) using the relation given in Eq.~(\ref{eq:relation_tau_L}). This shows that our model matches with the experimental observation that the optimal values of connecting tube length for achieving amplitude death are odd multiples of the length of the Rijke tube.

In Fig.~\ref{fig:twoparam}(b), we track the variation of the RMS value of the limit cycle oscillations ($p'_\text{rms}$) in the self-coupled Rijke tube over a small range of coupling strength for a fixed value of coupling delay ($\tau = 1$). We observe that the transition from limit cycle oscillatory state to amplitude death occurs through fold bifurcation on increasing $K_\tau$ in the forward path, while the transition back to the state of limit cycle oscillations occurs through subcritical Hopf bifurcation in the reverse path. As a result, we observe hysteresis or bistability between the states of limit cycle oscillations and amplitude death in the self-coupled Rijke tube. The bistable region surrounds each island of amplitude death, as seen in Fig.~\ref{fig:twoparam}(a). The bistable zone is significantly wider for higher values of coupling strength. The boundary demarcating the regions of amplitude death derived using linear stability analysis [obtained from Eq.~(\ref{eq11}) and (\ref{eq12})] matches well with the outer boundary of the bistable zone. This is because linear stability analysis predicts the local stability of the fixed point $\eta=0$, and thus finds the Hopf point of the system \cite{lakshmanan2011dynamics}. Slight differences exist between the analytical and numerical results due to our assumption of single mode.  


We next examine how variation in the non-dimensional heater power influences the behavior of the self-coupled Rijke tube. Figure \ref{fig:twoparam}(c) demonstrates the effect of the non-dimensional heater power $K$ on the acoustic pressure fluctuations in the system as $\tau$ is varied for a particular coupling strength ($K_\tau$ = 0.05). As we know, an increase in $K$ corresponds to an increase in the amplitude of limit cycle oscillations in the uncoupled state. We observe amplitude death for lower values of $K$, i.e., for low amplitudes of limit cycle oscillations in the uncoupled state. For sufficiently high values of $K$, i.e., high amplitudes of limit cycle oscillations in the uncoupled state, the oscillations continue to exist even on introducing self-coupling, regardless of the value of $\tau$. Furthermore, we observe that the boundary of the region of amplitude death oscillates with $\tau$, with the peaks at around $\tau = 1,\ 3,\ 5,\ ...$. On increasing $K$ at these values of $\tau$, the amplitude death region tapers out and subsequently disappears at high values. This behavior of the model qualitatively resembles the experimental results presented in Fig.~\ref{fig:p0_L}(b). We also note that the transition between the states of limit cycle oscillations and amplitude death on varying $K$ is explosive and hysteretic in Fig.~\ref{fig:twoparam}(d), similar to what we observe in experiments [refer to Fig.~\ref{fig2}(b)]. In Fig.~\ref{fig:twoparam}(c), we see that the boundary of the amplitude death region obtained using the method of averaging [from Eq.~(\ref{eq13})] matches well with the numerical results, especially for small values of $\tau$ where our simplifying assumptions are valid. 

\begin{figure}[t]
\includegraphics[width= 8.2cm]{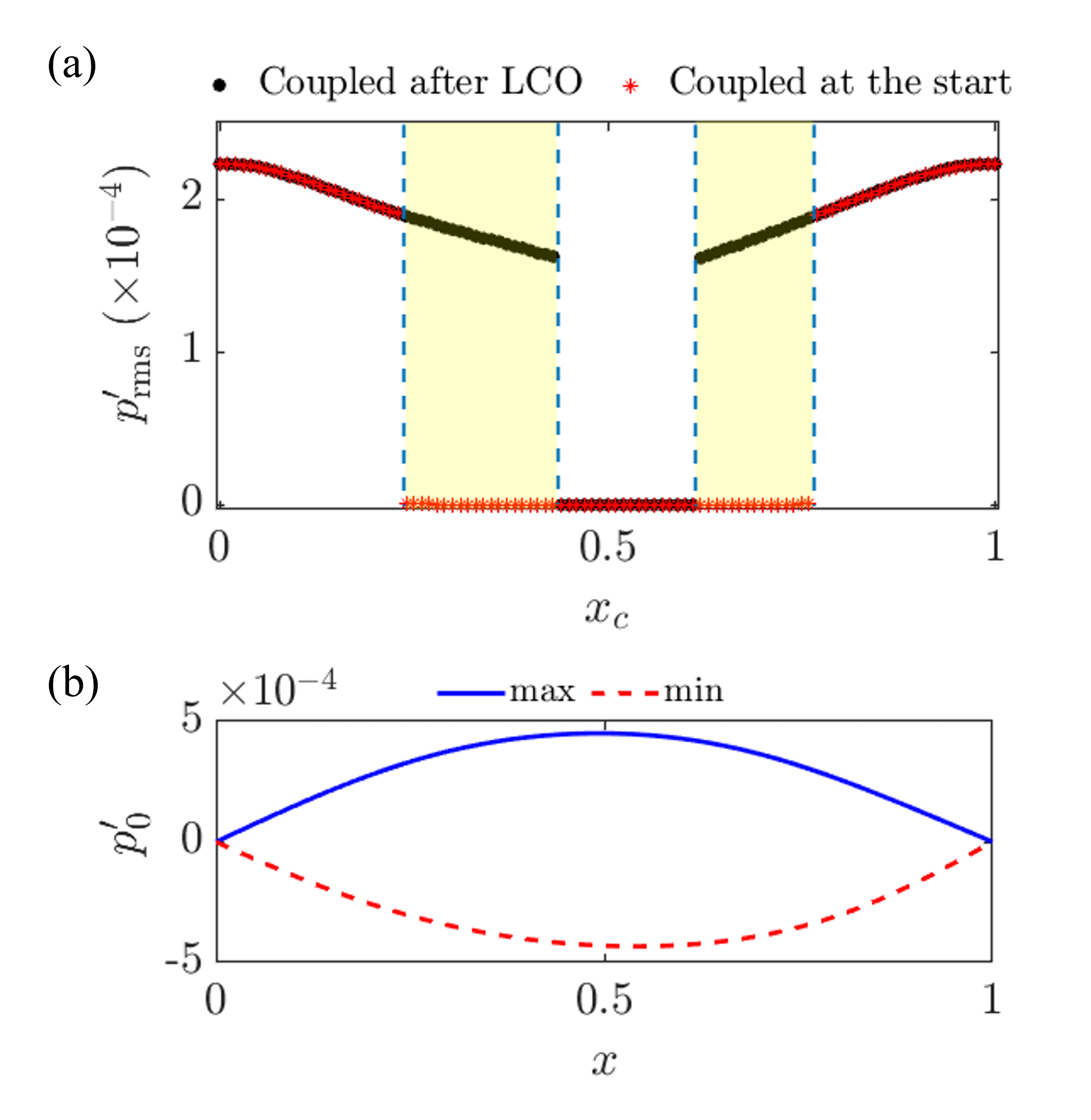}
\caption{\label{fig:xc}
(a) One-parameter bifurcation plot depicting the variation in $p'_\text{rms}$ with the location of the self-coupling tube on the duct  ($x_c$) for the self-coupled Rijke tube. The black (red) curve denotes the case when the Rijke tube is coupled after (before) it reaches limit cycle oscillatory state. The region highlighted in yellow denotes the values of $x_c$ for which amplitude death is achieved if the Rijke tube is coupled at the start but not on coupling after oscillations are established. (b) Standing wave pattern obtained by plotting the maximum and minimum values of the acoustic pressure oscillations ($p'_0$) in the uncoupled Rijke tube at different positions ($x$) along the duct. The antinode of the acoustic standing wave is observed at the middle of the duct, which is the optimal coupling location for achieving amplitude death in the self-coupled Rijke tube, as seen in (a). $K_\tau = 0.05$, $\tau = 1$, and $K=0.8$ are maintained in both the plots.}
\end{figure}

\begin{figure*}[t]
\centering
\includegraphics[width=13cm]{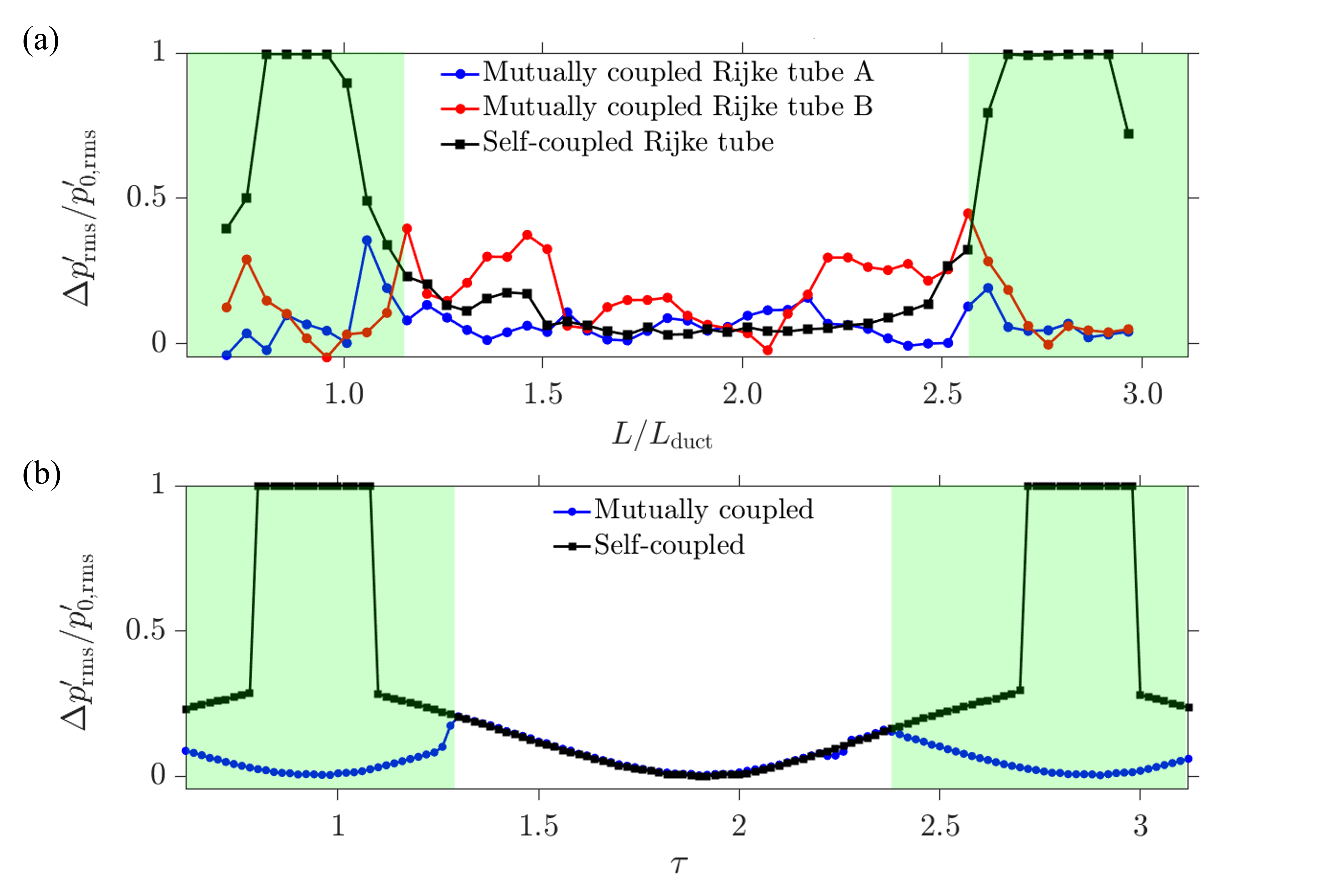}
\caption{\label{fig:comparison} Trends in the relative suppression of the acoustic pressure oscillations ($\Delta p'/p'_\text{0,rms}$) in two mutually coupled identical Rijke tubes and in a self-coupled Rijke tube with change in (a) the normalized length of the connecting tube ($L/L_\text{duct}$) in experiments ($d=8$ mm and $p'_\text{0,rms}=320$ Pa) and (b) the coupling delay ($\tau$) in the model ($K_\tau = 0.05$ and $K = 0.81$). The highlighted green region indicates the range of $L/L_\text{duct}$ or $\tau$ where the self-coupled Rijke tube exhibits greater amplitude suppression than the mutually coupled system.}
\end{figure*}

We now investigate the effect of the position of the coupling tube along the length of a Rijke tube on the suppression of limit cycle oscillations in the model. The position of the coupling tube is denoted as $x_c$ in the right-hand-side of Eq.~(\ref{eq9}). Figure \ref{fig:xc} illustrates the change in the RMS value of the limit cycle oscillations ($p'_\text{rms}$) on variation of $x_c$ while all other parameters are kept fixed in the system. We observe a region of amplitude death when the connecting tube is around midway along the duct in the model. On the other hand, when $x_c$ is towards the ends of the duct (i.e., $x_c$ is close to 0 or 1), we notice that the self-coupled Rijke tube retains the state of limit cycle oscillations. Thus, the model substantiates our observations from experiments (Fig.~\ref{fig:expt_loc} of Sec.~\ref{sec:IIIA}) where we noted the optimal coupling location for achieving amplitude death to be around the middle of the Rijke tube. 

In order to understand why the optimal value of $x_c$ leading to amplitude death is around the middle of the Rijke tube, we examine how the acoustic standing wave pattern looks in the Rijke tube prior to coupling [refer Fig.~\ref{fig:xc}(b)]. Since the first acoustic mode is primarily excited in the Rijke tube, the acoustic standing wave pattern has nodes at the end of the duct and an antinode at around the middle of the duct (at $x\approx 0.5$). Thus, on coupling the Rijke tube around the antinode of the acoustic standing wave, the Rijke tube receives the strongest acoustic feedback, easing the occurrence of AD. In Fig.~\ref{fig:xc}(a), we consider two cases of self-coupling: (i) when we let the Rijke tube attain limit cycle oscillations and then induce self-coupling, and (ii) when we couple the Rijke tube at the start when the transients are small, i.e., before limit cycle oscillations are established in the duct. We observe wider region of amplitude death in the second case when self-coupling is induced to begin with.


Thus, through experiments and modeling, we showed that self-coupling can bring about amplitude death of limit cycle oscillations in a thermoacoustic system for optimal values of coupling and system parameters. Recent studies have exploited the method of mutual coupling to quench limit cycle oscillations in two thermoacoustic oscillators \cite{thomas2018effect,dange2019oscillation,jegal2019mutual,hyodo2020suppression,sahay2021dynamics,srikanth2021dynamical}. Hence, we next compare how effective self-coupling is in quenching high amplitude limit cycle oscillations in a thermoacoustic oscillator as compared to mutual coupling of two such identical oscillators, through both experiments and modeling.

\subsection{\label{sec:IIIC}Comparison between self-coupled and mutually coupled Rijke tube oscillators}
 
\vspace{-4mm} In order to compare the amplitude suppression characteristics of limit cycle oscillations in a mutually coupled system of identical Rijke tubes and a single self-coupled Rijke tube, we set the same system and coupling parameters and measure the fractional change in the RMS value of limit cycle oscillations after coupling is induced in the system ($\Delta p'_\text{rms}/p'_\text{0,rms}$). We repeat this for different values of the normalized coupling tube length ($L/L_\text{duct}$) in experiments or the coupling delay ($\tau$) in the model. In both the experiments and the model [Fig.~\ref{fig:comparison}(a) and \ref{fig:comparison}(b), respectively], we observe that in the green regions present around odd values of $L/L_\text{duct}$ or $\tau$, the self-coupled system suppresses limit cycle oscillations to a greater extent than the mutually coupled system and even induces amplitude death. Away from these optimal values of $L/L_\text{duct}$ or $\tau$, the Rijke tube exhibits almost the same magnitude of amplitude suppression for both self and mutual couplings. Here, we note that previous studies have introduced frequency mismatch to quench high amplitude limit cycle oscillations in mutually coupled Rijke tubes \cite{dange2019oscillation,sahay2021dynamics}. We do not consider such mismatch in the current study. Thus, we emphasize that a self-coupled system performs significantly better in suppressing the oscillations in a nonlinear system, compared to the amplitude suppression in a mutually coupled system of identical oscillators. 


\section{\label{sec:IV}Conclusions}
In this study, we demonstrated that limit cycle oscillations can be quenched in a thermoacoustic system through the method of self-coupling, wherein we couple the acoustic field of a thermoacoustic oscillator to itself using a connecting tube. This method removes the need for electromechanical components that are present in traditional feedback control strategies. Through experiments and modeling, we found that the optimal lengths of the connecting tube to achieve amplitude death are around odd multiples of the length of the Rijke tube. Increasing the diameter of the connecting tube increases the ranges of the connecting tube length over which amplitude death can be achieved. Additionally, we observed that oscillations of low amplitude are more easily quenched as compared to those of high amplitude. Furthermore, our experiments and model indicate the optimal axial location of the connecting tube to be around the antinode of the acoustic standing wave in the uncoupled Rijke tube. We observed that the transition between steady state and oscillatory state is explosive and hysteretic in the self-coupled Rijke tube, similar to the behavior of the uncoupled Rijke tube. Furthermore, we demonstrated that self-coupling is more effective in achieving amplitude death in a Rijke tube as compared to mutual coupling of two such identical Rijke tubes. Thus, we anticipate self-coupling to be a simple, cost-effective alternative to traditional feedback controls for mitigating thermoacoustic instability in single and multiple thermoacoustic systems, such as those practical gas turbine and rocket combustors. It would be interesting to explore the mitigation of thermoacoustic instability through the combined application of self and mutual couplings in multiple combustion systems. Quenching transverse thermoacoustic instabilities in a combustor through self-coupling is also a scope for future studies.

\begin{acknowledgments}
 S. S. is grateful to Prof. Preeti Aghalayam and other members of the Young Research Fellow Program of Indian Institute of Technology Madras (Project ID: 202025), India. A. S. gratefully acknowledges the Half-Time Research Assistantship (HTRA) from the Ministry of Education. R. I. S. gratefully acknowledges the J. C. Bose Fellowship (No. JCB/2018/000034/SSC) from the Department of Science and Technology (DST), and the IoE initiative (SB/2021/0845/AE/MHRD/002696) for the financial support. We are thankful to Mr. Rohit R. for helping out with the experiments.
\end{acknowledgments}

\bibliography{apssamp}

\providecommand{\noopsort}[1]{}\providecommand{\singleletter}[1]{#1}%
\begin{thebibliography}{45}%
\makeatletter
\providecommand \@ifxundefined [1]{%
 \@ifx{#1\undefined}
}%
\providecommand \@ifnum [1]{%
 \ifnum #1\expandafter \@firstoftwo
 \else \expandafter \@secondoftwo
 \fi
}%
\providecommand \@ifx [1]{%
 \ifx #1\expandafter \@firstoftwo
 \else \expandafter \@secondoftwo
 \fi
}%
\providecommand \natexlab [1]{#1}%
\providecommand \enquote  [1]{``#1''}%
\providecommand \bibnamefont  [1]{#1}%
\providecommand \bibfnamefont [1]{#1}%
\providecommand \citenamefont [1]{#1}%
\providecommand \href@noop [0]{\@secondoftwo}%
\providecommand \href [0]{\begingroup \@sanitize@url \@href}%
\providecommand \@href[1]{\@@startlink{#1}\@@href}%
\providecommand \@@href[1]{\endgroup#1\@@endlink}%
\providecommand \@sanitize@url [0]{\catcode `\\12\catcode `\$12\catcode
  `\&12\catcode `\#12\catcode `\^12\catcode `\_12\catcode `\%12\relax}%
\providecommand \@@startlink[1]{}%
\providecommand \@@endlink[0]{}%
\providecommand \url  [0]{\begingroup\@sanitize@url \@url }%
\providecommand \@url [1]{\endgroup\@href {#1}{\urlprefix }}%
\providecommand \urlprefix  [0]{URL }%
\providecommand \Eprint [0]{\href }%
\providecommand \doibase [0]{https://doi.org/}%
\providecommand \selectlanguage [0]{\@gobble}%
\providecommand \bibinfo  [0]{\@secondoftwo}%
\providecommand \bibfield  [0]{\@secondoftwo}%
\providecommand \translation [1]{[#1]}%
\providecommand \BibitemOpen [0]{}%
\providecommand \bibitemStop [0]{}%
\providecommand \bibitemNoStop [0]{.\EOS\space}%
\providecommand \EOS [0]{\spacefactor3000\relax}%
\providecommand \BibitemShut  [1]{\csname bibitem#1\endcsname}%
\let\auto@bib@innerbib\@empty
\bibitem [{\citenamefont {Strogatz}\ \emph {et~al.}(2005)\citenamefont
  {Strogatz}, \citenamefont {Abrams}, \citenamefont {McRobie}, \citenamefont
  {Eckhardt},\ and\ \citenamefont {Ott}}]{strogatz2005crowd}%
  \BibitemOpen
  \bibfield  {author} {\bibinfo {author} {\bibfnamefont {S.~H.}\ \bibnamefont
  {Strogatz}}, \bibinfo {author} {\bibfnamefont {D.~M.}\ \bibnamefont
  {Abrams}}, \bibinfo {author} {\bibfnamefont {A.}~\bibnamefont {McRobie}},
  \bibinfo {author} {\bibfnamefont {B.}~\bibnamefont {Eckhardt}},\ and\
  \bibinfo {author} {\bibfnamefont {E.}~\bibnamefont {Ott}},\ }\bibfield
  {title} {\bibinfo {title} {Crowd {s}ynchrony on the {M}illennium {B}ridge},\
  }\href@noop {} {\bibfield  {journal} {\bibinfo  {journal} {Nature}\ }\textbf
  {\bibinfo {volume} {438}},\ \bibinfo {pages} {43} (\bibinfo {year}
  {2005})}\BibitemShut {NoStop}%
\bibitem [{\citenamefont {Garrick}\ and\ \citenamefont
  {Reed~III}(1981)}]{garrick1981historical}%
  \BibitemOpen
  \bibfield  {author} {\bibinfo {author} {\bibfnamefont {I.~E.}\ \bibnamefont
  {Garrick}}\ and\ \bibinfo {author} {\bibfnamefont {W.~H.}\ \bibnamefont
  {Reed~III}},\ }\bibfield  {title} {\bibinfo {title} {Historical {D}evelopment
  of {A}ircraft {F}lutter},\ }\href@noop {} {\bibfield  {journal} {\bibinfo
  {journal} {Journal of Aircraft}\ }\textbf {\bibinfo {volume} {18}},\ \bibinfo
  {pages} {897} (\bibinfo {year} {1981})}\BibitemShut {NoStop}%
\bibitem [{\citenamefont {Lieuwen}\ and\ \citenamefont
  {Yang}(2005)}]{lieuwen2005combustion}%
  \BibitemOpen
  \bibfield  {author} {\bibinfo {author} {\bibfnamefont {T.~C.}\ \bibnamefont
  {Lieuwen}}\ and\ \bibinfo {author} {\bibfnamefont {V.}~\bibnamefont {Yang}},\
  }\href@noop {} {\emph {\bibinfo {title} {Combustion {I}nstabilities in {G}as
  {T}urbine {E}ngines: {O}perational {E}xperience, {F}undamental {M}echanisms,
  and {M}odeling}}}\ (\bibinfo  {publisher} {American Institute of Aeronautics
  and Astronautics},\ \bibinfo {year} {2005})\BibitemShut {NoStop}%
\bibitem [{\citenamefont {Culick}(2006)}]{culick2006unsteady}%
  \BibitemOpen
  \bibfield  {author} {\bibinfo {author} {\bibfnamefont {F.~E.~C.}\
  \bibnamefont {Culick}},\ }\href@noop {} {\emph {\bibinfo {title} {Unsteady
  {M}otions in {C}ombustion {C}hambers for {P}ropulsion {S}ystems}}},\ \bibinfo
  {type} {Tech. Rep.}\ (\bibinfo  {institution} {AGARDograph,
  NATO/RTO-AG-AVT-039},\ \bibinfo {year} {2006})\BibitemShut {NoStop}%
\bibitem [{\citenamefont {May}(2019)}]{may2019stability}%
  \BibitemOpen
  \bibfield  {author} {\bibinfo {author} {\bibfnamefont {R.~M.}\ \bibnamefont
  {May}},\ }\href@noop {} {\emph {\bibinfo {title} {Stability and {C}omplexity
  in {M}odel {E}cosystems}}}\ (\bibinfo  {publisher} {Princeton university
  press},\ \bibinfo {year} {2019})\BibitemShut {NoStop}%
\bibitem [{\citenamefont {Zou}\ \emph {et~al.}(2021)\citenamefont {Zou},
  \citenamefont {Senthilkumar}, \citenamefont {Zhan},\ and\ \citenamefont
  {Kurths}}]{zou2021quenching}%
  \BibitemOpen
  \bibfield  {author} {\bibinfo {author} {\bibfnamefont {W.}~\bibnamefont
  {Zou}}, \bibinfo {author} {\bibfnamefont {D.~V.}\ \bibnamefont
  {Senthilkumar}}, \bibinfo {author} {\bibfnamefont {M.}~\bibnamefont {Zhan}},\
  and\ \bibinfo {author} {\bibfnamefont {J.}~\bibnamefont {Kurths}},\
  }\bibfield  {title} {\bibinfo {title} {Quenching, aging, and reviving in
  coupled dynamical networks},\ }\href@noop {} {\bibfield  {journal} {\bibinfo
  {journal} {Physics Reports}\ } (\bibinfo {year} {2021})}\BibitemShut
  {NoStop}%
\bibitem [{\citenamefont {Frankel}(2008)}]{frankel2008adaptive}%
  \BibitemOpen
  \bibfield  {author} {\bibinfo {author} {\bibfnamefont {D.~M.}\ \bibnamefont
  {Frankel}},\ }\bibfield  {title} {\bibinfo {title} {Adaptive expectations and
  stock market crashes},\ }\href@noop {} {\bibfield  {journal} {\bibinfo
  {journal} {International Economic Review}\ }\textbf {\bibinfo {volume}
  {49}},\ \bibinfo {pages} {595} (\bibinfo {year} {2008})}\BibitemShut
  {NoStop}%
\bibitem [{\citenamefont {Sujith}\ and\ \citenamefont
  {Unni}(2020)}]{sujith2020complex}%
  \BibitemOpen
  \bibfield  {author} {\bibinfo {author} {\bibfnamefont {R.~I.}\ \bibnamefont
  {Sujith}}\ and\ \bibinfo {author} {\bibfnamefont {V.~R.}\ \bibnamefont
  {Unni}},\ }\bibfield  {title} {\bibinfo {title} {Complex system approach to
  investigate and mitigate thermoacoustic instability in turbulent
  combustors},\ }\href@noop {} {\bibfield  {journal} {\bibinfo  {journal}
  {Physics of Fluids}\ }\textbf {\bibinfo {volume} {32}},\ \bibinfo {pages}
  {061401} (\bibinfo {year} {2020})}\BibitemShut {NoStop}%
\bibitem [{\citenamefont {Poinsot}(2017)}]{poinsot2017prediction}%
  \BibitemOpen
  \bibfield  {author} {\bibinfo {author} {\bibfnamefont {T.}~\bibnamefont
  {Poinsot}},\ }\bibfield  {title} {\bibinfo {title} {Prediction and control of
  combustion instabilities in real engines},\ }\href@noop {} {\bibfield
  {journal} {\bibinfo  {journal} {Proceedings of the Combustion Institute}\
  }\textbf {\bibinfo {volume} {36}},\ \bibinfo {pages} {1} (\bibinfo {year}
  {2017})}\BibitemShut {NoStop}%
\bibitem [{\citenamefont {McManus}\ \emph {et~al.}(1993)\citenamefont
  {McManus}, \citenamefont {Poinsot},\ and\ \citenamefont
  {Candel}}]{mcmanus1993review}%
  \BibitemOpen
  \bibfield  {author} {\bibinfo {author} {\bibfnamefont {K.~R.}\ \bibnamefont
  {McManus}}, \bibinfo {author} {\bibfnamefont {T.}~\bibnamefont {Poinsot}},\
  and\ \bibinfo {author} {\bibfnamefont {S.~M.}\ \bibnamefont {Candel}},\
  }\bibfield  {title} {\bibinfo {title} {A review of active control of
  combustion instabilities},\ }\href@noop {} {\bibfield  {journal} {\bibinfo
  {journal} {Progress in Energy and Combustion Science}\ }\textbf {\bibinfo
  {volume} {19}},\ \bibinfo {pages} {1} (\bibinfo {year} {1993})}\BibitemShut
  {NoStop}%
\bibitem [{\citenamefont {Dowling}\ and\ \citenamefont
  {Morgans}(2005)}]{dowling2005feedback}%
  \BibitemOpen
  \bibfield  {author} {\bibinfo {author} {\bibfnamefont {A.~P.}\ \bibnamefont
  {Dowling}}\ and\ \bibinfo {author} {\bibfnamefont {A.~S.}\ \bibnamefont
  {Morgans}},\ }\bibfield  {title} {\bibinfo {title} {Feedback {C}ontrol of
  {C}ombustion {O}scillations},\ }\href@noop {} {\bibfield  {journal} {\bibinfo
   {journal} {Annual Review of Fluid Mechanics}\ }\textbf {\bibinfo {volume}
  {37}},\ \bibinfo {pages} {151} (\bibinfo {year} {2005})}\BibitemShut
  {NoStop}%
\bibitem [{\citenamefont {Sujith}\ and\ \citenamefont
  {Pawar}(2021)}]{sujith2021book}%
  \BibitemOpen
  \bibfield  {author} {\bibinfo {author} {\bibfnamefont {R.~I.}\ \bibnamefont
  {Sujith}}\ and\ \bibinfo {author} {\bibfnamefont {S.~A.}\ \bibnamefont
  {Pawar}},\ }\href@noop {} {\emph {\bibinfo {title} {Thermoacoustic
  {I}nstability: {A} {C}omplex {S}ystems {P}erspective}}}\ (\bibinfo
  {publisher} {Springer International Publishing},\ \bibinfo {year}
  {2021})\BibitemShut {NoStop}%
\bibitem [{\citenamefont {Putnam}(1971)}]{putnam1971combustion}%
  \BibitemOpen
  \bibfield  {author} {\bibinfo {author} {\bibfnamefont {A.~A.}\ \bibnamefont
  {Putnam}},\ }\href@noop {} {\emph {\bibinfo {title} {Combustion-{D}riven
  {O}scillations in {I}ndustry}}}\ (\bibinfo  {publisher} {Elsevier},\ \bibinfo
  {year} {1971})\BibitemShut {NoStop}%
\bibitem [{\citenamefont {{\'C}osi{\'c}}\ \emph {et~al.}(2012)\citenamefont
  {{\'C}osi{\'c}}, \citenamefont {Bobusch}, \citenamefont {Moeck},\ and\
  \citenamefont {Paschereit}}]{cosic2012open}%
  \BibitemOpen
  \bibfield  {author} {\bibinfo {author} {\bibfnamefont {B.}~\bibnamefont
  {{\'C}osi{\'c}}}, \bibinfo {author} {\bibfnamefont {B.~C.}\ \bibnamefont
  {Bobusch}}, \bibinfo {author} {\bibfnamefont {J.~P.}\ \bibnamefont {Moeck}},\
  and\ \bibinfo {author} {\bibfnamefont {C.~O.}\ \bibnamefont {Paschereit}},\
  }\bibfield  {title} {\bibinfo {title} {Open-{L}oop {C}ontrol of {C}ombustion
  {I}nstabilities and the {R}ole of the {F}lame {R}esponse to {T}wo-{F}requency
  {F}orcing},\ }\href@noop {} {\bibfield  {journal} {\bibinfo  {journal}
  {Journal of engineering for gas turbines and power}\ }\textbf {\bibinfo
  {volume} {134}} (\bibinfo {year} {2012})}\BibitemShut {NoStop}%
\bibitem [{\citenamefont {Heckl}(1988)}]{heckl1988active}%
  \BibitemOpen
  \bibfield  {author} {\bibinfo {author} {\bibfnamefont {M.~A.}\ \bibnamefont
  {Heckl}},\ }\bibfield  {title} {\bibinfo {title} {Active control of the noise
  from a {R}ijke tube},\ }\href@noop {} {\bibfield  {journal} {\bibinfo
  {journal} {Journal of Sound and Vibration}\ }\textbf {\bibinfo {volume}
  {124}},\ \bibinfo {pages} {117} (\bibinfo {year} {1988})}\BibitemShut
  {NoStop}%
\bibitem [{\citenamefont {Neumeier}\ and\ \citenamefont
  {Zinn}(1996)}]{neumeier1996active}%
  \BibitemOpen
  \bibfield  {author} {\bibinfo {author} {\bibfnamefont {Y.}~\bibnamefont
  {Neumeier}}\ and\ \bibinfo {author} {\bibfnamefont {B.}~\bibnamefont
  {Zinn}},\ }\bibfield  {title} {\bibinfo {title} {Active control of combustion
  instabilities with real time operation of unstable combustor modes},\ }in\
  \href@noop {} {\emph {\bibinfo {booktitle} {34th Aerospace Sciences Meeting
  and Exhibit}}}\ (\bibinfo {year} {1996})\ p.\ \bibinfo {pages}
  {758}\BibitemShut {NoStop}%
\bibitem [{\citenamefont {Annaswamy}\ and\ \citenamefont
  {Ghoniem}(1995)}]{annaswamy1995active}%
  \BibitemOpen
  \bibfield  {author} {\bibinfo {author} {\bibfnamefont {A.~M.}\ \bibnamefont
  {Annaswamy}}\ and\ \bibinfo {author} {\bibfnamefont {A.~F.}\ \bibnamefont
  {Ghoniem}},\ }\bibfield  {title} {\bibinfo {title} {Active control in
  combustion systems},\ }\href@noop {} {\bibfield  {journal} {\bibinfo
  {journal} {IEEE Control Systems Magazine}\ }\textbf {\bibinfo {volume}
  {15}},\ \bibinfo {pages} {49} (\bibinfo {year} {1995})}\BibitemShut {NoStop}%
\bibitem [{\citenamefont {Mirollo}\ and\ \citenamefont
  {Strogatz}(1990)}]{mirollo1990amplitude}%
  \BibitemOpen
  \bibfield  {author} {\bibinfo {author} {\bibfnamefont {R.~E.}\ \bibnamefont
  {Mirollo}}\ and\ \bibinfo {author} {\bibfnamefont {S.~H.}\ \bibnamefont
  {Strogatz}},\ }\bibfield  {title} {\bibinfo {title} {Amplitude death in an
  array of limit-cycle oscillators},\ }\href@noop {} {\bibfield  {journal}
  {\bibinfo  {journal} {Journal of Statistical Physics}\ }\textbf {\bibinfo
  {volume} {60}},\ \bibinfo {pages} {245} (\bibinfo {year} {1990})}\BibitemShut
  {NoStop}%
\bibitem [{\citenamefont {Reddy}\ \emph {et~al.}(2000)\citenamefont {Reddy},
  \citenamefont {Sen},\ and\ \citenamefont {Johnston}}]{reddy2000dynamics}%
  \BibitemOpen
  \bibfield  {author} {\bibinfo {author} {\bibfnamefont {D.~V.~R.}\
  \bibnamefont {Reddy}}, \bibinfo {author} {\bibfnamefont {A.}~\bibnamefont
  {Sen}},\ and\ \bibinfo {author} {\bibfnamefont {G.~L.}\ \bibnamefont
  {Johnston}},\ }\bibfield  {title} {\bibinfo {title} {Dynamics of a {L}imit
  {C}ycle {O}scillator under {T}ime {D}elayed {L}inear and {N}onlinear
  {F}eedbacks},\ }\href@noop {} {\bibfield  {journal} {\bibinfo  {journal}
  {Physica D: Nonlinear Phenomena}\ }\textbf {\bibinfo {volume} {144}},\
  \bibinfo {pages} {335} (\bibinfo {year} {2000})}\BibitemShut {NoStop}%
\bibitem [{\citenamefont {Biwa}\ \emph {et~al.}(2016)\citenamefont {Biwa},
  \citenamefont {Sawada}, \citenamefont {Hyodo},\ and\ \citenamefont
  {Kato}}]{biwa2016suppression}%
  \BibitemOpen
  \bibfield  {author} {\bibinfo {author} {\bibfnamefont {T.}~\bibnamefont
  {Biwa}}, \bibinfo {author} {\bibfnamefont {Y.}~\bibnamefont {Sawada}},
  \bibinfo {author} {\bibfnamefont {H.}~\bibnamefont {Hyodo}},\ and\ \bibinfo
  {author} {\bibfnamefont {S.}~\bibnamefont {Kato}},\ }\bibfield  {title}
  {\bibinfo {title} {Suppression of {S}pontaneous {G}as {O}scillations by
  {A}coustic {S}elf-{F}eedback},\ }\href@noop {} {\bibfield  {journal}
  {\bibinfo  {journal} {Physical Review Applied}\ }\textbf {\bibinfo {volume}
  {6}},\ \bibinfo {pages} {044020} (\bibinfo {year} {2016})}\BibitemShut
  {NoStop}%
\bibitem [{\citenamefont {Suchorsky}\ \emph {et~al.}(2010)\citenamefont
  {Suchorsky}, \citenamefont {Sah},\ and\ \citenamefont
  {Rand}}]{suchorsky2010using}%
  \BibitemOpen
  \bibfield  {author} {\bibinfo {author} {\bibfnamefont {M.~K.}\ \bibnamefont
  {Suchorsky}}, \bibinfo {author} {\bibfnamefont {S.~M.}\ \bibnamefont {Sah}},\
  and\ \bibinfo {author} {\bibfnamefont {R.~H.}\ \bibnamefont {Rand}},\
  }\bibfield  {title} {\bibinfo {title} {Using delay to quench undesirable
  vibrations},\ }\href@noop {} {\bibfield  {journal} {\bibinfo  {journal}
  {Nonlinear Dynamics}\ }\textbf {\bibinfo {volume} {62}},\ \bibinfo {pages}
  {407} (\bibinfo {year} {2010})}\BibitemShut {NoStop}%
\bibitem [{\citenamefont {Ahlborn}\ and\ \citenamefont
  {Parlitz}(2005)}]{ahlborn2005controlling}%
  \BibitemOpen
  \bibfield  {author} {\bibinfo {author} {\bibfnamefont {A.}~\bibnamefont
  {Ahlborn}}\ and\ \bibinfo {author} {\bibfnamefont {U.}~\bibnamefont
  {Parlitz}},\ }\bibfield  {title} {\bibinfo {title} {Controlling dynamical
  systems using multiple delay feedback control},\ }\href@noop {} {\bibfield
  {journal} {\bibinfo  {journal} {Physical Review E}\ }\textbf {\bibinfo
  {volume} {72}},\ \bibinfo {pages} {016206} (\bibinfo {year}
  {2005})}\BibitemShut {NoStop}%
\bibitem [{\citenamefont {Parmananda}\ \emph {et~al.}(1999)\citenamefont
  {Parmananda}, \citenamefont {Madrigal}, \citenamefont {Rivera}, \citenamefont
  {Nyikos}, \citenamefont {Kiss},\ and\ \citenamefont
  {G{\'a}sp{\'a}r}}]{parmananda1999stabilization}%
  \BibitemOpen
  \bibfield  {author} {\bibinfo {author} {\bibfnamefont {P.}~\bibnamefont
  {Parmananda}}, \bibinfo {author} {\bibfnamefont {R.}~\bibnamefont
  {Madrigal}}, \bibinfo {author} {\bibfnamefont {M.}~\bibnamefont {Rivera}},
  \bibinfo {author} {\bibfnamefont {L.}~\bibnamefont {Nyikos}}, \bibinfo
  {author} {\bibfnamefont {I.~Z.}\ \bibnamefont {Kiss}},\ and\ \bibinfo
  {author} {\bibfnamefont {V.}~\bibnamefont {G{\'a}sp{\'a}r}},\ }\bibfield
  {title} {\bibinfo {title} {Stabilization of unstable steady states and
  periodic orbits in an electrochemical system using delayed-feedback
  control},\ }\href@noop {} {\bibfield  {journal} {\bibinfo  {journal}
  {Physical Review E}\ }\textbf {\bibinfo {volume} {59}},\ \bibinfo {pages}
  {5266} (\bibinfo {year} {1999})}\BibitemShut {NoStop}%
\bibitem [{\citenamefont {Naumann}\ \emph {et~al.}(2014)\citenamefont
  {Naumann}, \citenamefont {Hein}, \citenamefont {Knorr},\ and\ \citenamefont
  {Kabuss}}]{naumann2014steady}%
  \BibitemOpen
  \bibfield  {author} {\bibinfo {author} {\bibfnamefont {N.~L.}\ \bibnamefont
  {Naumann}}, \bibinfo {author} {\bibfnamefont {S.~M.}\ \bibnamefont {Hein}},
  \bibinfo {author} {\bibfnamefont {A.}~\bibnamefont {Knorr}},\ and\ \bibinfo
  {author} {\bibfnamefont {J.}~\bibnamefont {Kabuss}},\ }\bibfield  {title}
  {\bibinfo {title} {Steady-state control in an unstable optomechanical
  system},\ }\href@noop {} {\bibfield  {journal} {\bibinfo  {journal} {Physical
  Review A}\ }\textbf {\bibinfo {volume} {90}},\ \bibinfo {pages} {043835}
  (\bibinfo {year} {2014})}\BibitemShut {NoStop}%
\bibitem [{\citenamefont {Lato}\ \emph {et~al.}(2019)\citenamefont {Lato},
  \citenamefont {Mohany},\ and\ \citenamefont {Hassan}}]{lato2019passive}%
  \BibitemOpen
  \bibfield  {author} {\bibinfo {author} {\bibfnamefont {T.}~\bibnamefont
  {Lato}}, \bibinfo {author} {\bibfnamefont {A.}~\bibnamefont {Mohany}},\ and\
  \bibinfo {author} {\bibfnamefont {M.}~\bibnamefont {Hassan}},\ }\bibfield
  {title} {\bibinfo {title} {A passive damping device for suppressing acoustic
  pressure pulsations: {T}he infinity tube},\ }\href@noop {} {\bibfield
  {journal} {\bibinfo  {journal} {The Journal of the Acoustical Society of
  America}\ }\textbf {\bibinfo {volume} {146}},\ \bibinfo {pages} {4534}
  (\bibinfo {year} {2019})}\BibitemShut {NoStop}%
\bibitem [{\citenamefont {Rijke}(1859)}]{rijke1859lxxi}%
  \BibitemOpen
  \bibfield  {author} {\bibinfo {author} {\bibfnamefont {P.~L.}\ \bibnamefont
  {Rijke}},\ }\bibfield  {title} {\bibinfo {title} {On the {V}ibration of the
  {A}ir in a {R}ijke {T}ube {O}pen at {B}oth {E}nds},\ }\href@noop {}
  {\bibfield  {journal} {\bibinfo  {journal} {Philosophical Magazine}\ }\textbf
  {\bibinfo {volume} {17}},\ \bibinfo {pages} {419} (\bibinfo {year}
  {1859})}\BibitemShut {NoStop}%
\bibitem [{\citenamefont {Raun}\ \emph {et~al.}(1993)\citenamefont {Raun},
  \citenamefont {Beckstead}, \citenamefont {Finlinson},\ and\ \citenamefont
  {Brooks}}]{raun1993review}%
  \BibitemOpen
  \bibfield  {author} {\bibinfo {author} {\bibfnamefont {R.~L.}\ \bibnamefont
  {Raun}}, \bibinfo {author} {\bibfnamefont {M.~W.}\ \bibnamefont {Beckstead}},
  \bibinfo {author} {\bibfnamefont {J.~C.}\ \bibnamefont {Finlinson}},\ and\
  \bibinfo {author} {\bibfnamefont {K.~P.}\ \bibnamefont {Brooks}},\ }\bibfield
   {title} {\bibinfo {title} {A review of {R}ijke tubes, {R}ijke burners and
  related devices},\ }\href@noop {} {\bibfield  {journal} {\bibinfo  {journal}
  {Progress in Energy and Combustion science}\ }\textbf {\bibinfo {volume}
  {19}},\ \bibinfo {pages} {313} (\bibinfo {year} {1993})}\BibitemShut
  {NoStop}%
\bibitem [{\citenamefont {Matveev}(2003)}]{matveev2003thermoacoustic}%
  \BibitemOpen
  \bibfield  {author} {\bibinfo {author} {\bibfnamefont {K.~I.}\ \bibnamefont
  {Matveev}},\ }\href@noop {} {\emph {\bibinfo {title} {Thermoacoustic
  instabilities in the {R}ijke tube: {E}xperiments and modeling}}}\ (\bibinfo
  {publisher} {California Institute of Technology},\ \bibinfo {year}
  {2003})\BibitemShut {NoStop}%
\bibitem [{\citenamefont {Dange}\ \emph {et~al.}(2019)\citenamefont {Dange},
  \citenamefont {Manoj}, \citenamefont {Banerjee}, \citenamefont {Pawar},
  \citenamefont {Mondal},\ and\ \citenamefont {Sujith}}]{dange2019oscillation}%
  \BibitemOpen
  \bibfield  {author} {\bibinfo {author} {\bibfnamefont {S.}~\bibnamefont
  {Dange}}, \bibinfo {author} {\bibfnamefont {K.}~\bibnamefont {Manoj}},
  \bibinfo {author} {\bibfnamefont {S.}~\bibnamefont {Banerjee}}, \bibinfo
  {author} {\bibfnamefont {S.~A.}\ \bibnamefont {Pawar}}, \bibinfo {author}
  {\bibfnamefont {S.}~\bibnamefont {Mondal}},\ and\ \bibinfo {author}
  {\bibfnamefont {R.~I.}\ \bibnamefont {Sujith}},\ }\bibfield  {title}
  {\bibinfo {title} {Oscillation quenching and phase-flip bifurcation in
  coupled thermoacoustic systems},\ }\href@noop {} {\bibfield  {journal}
  {\bibinfo  {journal} {Chaos: An Interdisciplinary Journal of Nonlinear
  Science}\ }\textbf {\bibinfo {volume} {29}},\ \bibinfo {pages} {093135}
  (\bibinfo {year} {2019})}\BibitemShut {NoStop}%
\bibitem [{\citenamefont {Sahay}\ \emph {et~al.}(2021)\citenamefont {Sahay},
  \citenamefont {Roy}, \citenamefont {Pawar},\ and\ \citenamefont
  {Sujith}}]{sahay2021dynamics}%
  \BibitemOpen
  \bibfield  {author} {\bibinfo {author} {\bibfnamefont {A.}~\bibnamefont
  {Sahay}}, \bibinfo {author} {\bibfnamefont {A.}~\bibnamefont {Roy}}, \bibinfo
  {author} {\bibfnamefont {S.~A.}\ \bibnamefont {Pawar}},\ and\ \bibinfo
  {author} {\bibfnamefont {R.~I.}\ \bibnamefont {Sujith}},\ }\bibfield  {title}
  {\bibinfo {title} {Dynamics of coupled thermoacoustic oscillators under
  asymmetric forcing},\ }\href@noop {} {\bibfield  {journal} {\bibinfo
  {journal} {Physical Review Applied}\ }\textbf {\bibinfo {volume} {15}},\
  \bibinfo {pages} {044011} (\bibinfo {year} {2021})}\BibitemShut {NoStop}%
\bibitem [{\citenamefont {Hyodo}\ \emph {et~al.}(2020)\citenamefont {Hyodo},
  \citenamefont {Iwasaki},\ and\ \citenamefont {Biwa}}]{hyodo2020suppression}%
  \BibitemOpen
  \bibfield  {author} {\bibinfo {author} {\bibfnamefont {H.}~\bibnamefont
  {Hyodo}}, \bibinfo {author} {\bibfnamefont {M.}~\bibnamefont {Iwasaki}},\
  and\ \bibinfo {author} {\bibfnamefont {T.}~\bibnamefont {Biwa}},\ }\bibfield
  {title} {\bibinfo {title} {Suppression of {R}ijke tube oscillations by delay
  coupling},\ }\href@noop {} {\bibfield  {journal} {\bibinfo  {journal}
  {Journal of Applied Physics}\ }\textbf {\bibinfo {volume} {128}},\ \bibinfo
  {pages} {094902} (\bibinfo {year} {2020})}\BibitemShut {NoStop}%
\bibitem [{\citenamefont {Srikanth}\ \emph {et~al.}(2021)\citenamefont
  {Srikanth}, \citenamefont {Pawar}, \citenamefont {Manoj},\ and\ \citenamefont
  {Sujith}}]{srikanth2021dynamical}%
  \BibitemOpen
  \bibfield  {author} {\bibinfo {author} {\bibfnamefont {S.}~\bibnamefont
  {Srikanth}}, \bibinfo {author} {\bibfnamefont {S.~A.}\ \bibnamefont {Pawar}},
  \bibinfo {author} {\bibfnamefont {K.}~\bibnamefont {Manoj}},\ and\ \bibinfo
  {author} {\bibfnamefont {R.~I.}\ \bibnamefont {Sujith}},\ }\href@noop {}
  {\bibinfo {title} {{D}ynamical {S}tates and {B}ifurcations in {C}oupled
  {T}hermoacoustic {O}scillators}} (\bibinfo {year} {2021}),\ \Eprint
  {https://arxiv.org/abs/2109.09600} {arXiv:2109.09600 [nlin.AO]} \BibitemShut
  {NoStop}%
\bibitem [{\citenamefont {Gopalakrishnan}\ and\ \citenamefont
  {Sujith}(2014)}]{gopalakrishnan2014influence}%
  \BibitemOpen
  \bibfield  {author} {\bibinfo {author} {\bibfnamefont {E.~A.}\ \bibnamefont
  {Gopalakrishnan}}\ and\ \bibinfo {author} {\bibfnamefont {R.~I.}\
  \bibnamefont {Sujith}},\ }\bibfield  {title} {\bibinfo {title} {Influence of
  system parameters on the hysteresis characteristics of a horizontal {R}ijke
  tube},\ }\href@noop {} {\bibfield  {journal} {\bibinfo  {journal}
  {International Journal of Spray and Combustion Dynamics}\ }\textbf {\bibinfo
  {volume} {6}},\ \bibinfo {pages} {293} (\bibinfo {year} {2014})}\BibitemShut
  {NoStop}%
\bibitem [{\citenamefont {Etikyala}\ and\ \citenamefont
  {Sujith}(2017)}]{etikyala2017change}%
  \BibitemOpen
  \bibfield  {author} {\bibinfo {author} {\bibfnamefont {S.}~\bibnamefont
  {Etikyala}}\ and\ \bibinfo {author} {\bibfnamefont {R.~I.}\ \bibnamefont
  {Sujith}},\ }\bibfield  {title} {\bibinfo {title} {Change of criticality in a
  prototypical thermoacoustic system},\ }\href@noop {} {\bibfield  {journal}
  {\bibinfo  {journal} {Chaos: An Interdisciplinary Journal of Nonlinear
  Science}\ }\textbf {\bibinfo {volume} {27}},\ \bibinfo {pages} {023106}
  (\bibinfo {year} {2017})}\BibitemShut {NoStop}%
\bibitem [{\citenamefont {Balasubramanian}\ and\ \citenamefont
  {Sujith}(2008)}]{balasubramanian2008thermoacoustic}%
  \BibitemOpen
  \bibfield  {author} {\bibinfo {author} {\bibfnamefont {K.}~\bibnamefont
  {Balasubramanian}}\ and\ \bibinfo {author} {\bibfnamefont {R.~I.}\
  \bibnamefont {Sujith}},\ }\bibfield  {title} {\bibinfo {title}
  {Thermoacoustic instability in a {R}ijke tube: {N}on-normality and
  nonlinearity},\ }\href@noop {} {\bibfield  {journal} {\bibinfo  {journal}
  {Physics of Fluids}\ }\textbf {\bibinfo {volume} {20}},\ \bibinfo {pages}
  {044103} (\bibinfo {year} {2008})}\BibitemShut {NoStop}%
\bibitem [{\citenamefont {Nicoud}\ and\ \citenamefont
  {Wieczorek}(2009)}]{nicoud2009zero}%
  \BibitemOpen
  \bibfield  {author} {\bibinfo {author} {\bibfnamefont {F.}~\bibnamefont
  {Nicoud}}\ and\ \bibinfo {author} {\bibfnamefont {K.}~\bibnamefont
  {Wieczorek}},\ }\bibfield  {title} {\bibinfo {title} {About the zero {M}ach
  number assumption in the calculation of thermoacoustic instabilities},\
  }\href@noop {} {\bibfield  {journal} {\bibinfo  {journal} {International
  Journal of Spray and Combustion Dynamics}\ }\textbf {\bibinfo {volume} {1}},\
  \bibinfo {pages} {67} (\bibinfo {year} {2009})}\BibitemShut {NoStop}%
\bibitem [{\citenamefont {Heckl}(1990)}]{heckl1990non}%
  \BibitemOpen
  \bibfield  {author} {\bibinfo {author} {\bibfnamefont {M.~A.}\ \bibnamefont
  {Heckl}},\ }\bibfield  {title} {\bibinfo {title} {Non-linear acoustic effects
  in the {R}ijke tube},\ }\href@noop {} {\bibfield  {journal} {\bibinfo
  {journal} {Acta Acustica united with Acustica}\ }\textbf {\bibinfo {volume}
  {72}},\ \bibinfo {pages} {63} (\bibinfo {year} {1990})}\BibitemShut {NoStop}%
\bibitem [{\citenamefont {Lighthill}(1954)}]{lighthill1954response}%
  \BibitemOpen
  \bibfield  {author} {\bibinfo {author} {\bibfnamefont {M.~J.}\ \bibnamefont
  {Lighthill}},\ }\bibfield  {title} {\bibinfo {title} {The response of laminar
  skin friction and heat transfer to fluctuations in the stream velocity},\
  }\href@noop {} {\bibfield  {journal} {\bibinfo  {journal} {Proceedings of the
  Royal Society of London. Series A. Mathematical and Physical Sciences}\
  }\textbf {\bibinfo {volume} {224}},\ \bibinfo {pages} {1} (\bibinfo {year}
  {1954})}\BibitemShut {NoStop}%
\bibitem [{\citenamefont {Lores}\ and\ \citenamefont
  {Zinn}(1973)}]{lores1973nonlinear}%
  \BibitemOpen
  \bibfield  {author} {\bibinfo {author} {\bibfnamefont {M.~E.}\ \bibnamefont
  {Lores}}\ and\ \bibinfo {author} {\bibfnamefont {B.~T.}\ \bibnamefont
  {Zinn}},\ }\bibfield  {title} {\bibinfo {title} {Nonlinear {L}ongitudinal
  {C}ombustion {I}nstability in {R}ocket {M}otors},\ }\href@noop {} {\bibfield
  {journal} {\bibinfo  {journal} {Combustion Science and Technology}\ }\textbf
  {\bibinfo {volume} {7}},\ \bibinfo {pages} {245} (\bibinfo {year}
  {1973})}\BibitemShut {NoStop}%
\bibitem [{\citenamefont {Subramanian}\ \emph {et~al.}(2013)\citenamefont
  {Subramanian}, \citenamefont {Sujith},\ and\ \citenamefont
  {Wahi}}]{subramanian2013subcritical}%
  \BibitemOpen
  \bibfield  {author} {\bibinfo {author} {\bibfnamefont {P.}~\bibnamefont
  {Subramanian}}, \bibinfo {author} {\bibfnamefont {R.~I.}\ \bibnamefont
  {Sujith}},\ and\ \bibinfo {author} {\bibfnamefont {P.}~\bibnamefont {Wahi}},\
  }\bibfield  {title} {\bibinfo {title} {Subcritical bifurcation and
  bistability in thermoacoustic systems},\ }\href@noop {} {\bibfield  {journal}
  {\bibinfo  {journal} {Journal of Fluid Mechanics}\ }\textbf {\bibinfo
  {volume} {715}},\ \bibinfo {pages} {210} (\bibinfo {year}
  {2013})}\BibitemShut {NoStop}%
\bibitem [{\citenamefont {Lakshmanan}\ and\ \citenamefont
  {Senthilkumar}(2011)}]{lakshmanan2011dynamics}%
  \BibitemOpen
  \bibfield  {author} {\bibinfo {author} {\bibfnamefont {M.}~\bibnamefont
  {Lakshmanan}}\ and\ \bibinfo {author} {\bibfnamefont {D.~V.}\ \bibnamefont
  {Senthilkumar}},\ }\href@noop {} {\emph {\bibinfo {title} {Dynamics of
  {N}onlinear {T}ime-{D}elay {S}ystems}}}\ (\bibinfo  {publisher} {Springer
  Science \& Business Media},\ \bibinfo {year} {2011})\BibitemShut {NoStop}%
\bibitem [{\citenamefont {Shampine}\ \emph {et~al.}(2000)\citenamefont
  {Shampine}, \citenamefont {Thompson},\ and\ \citenamefont
  {Kierzenka}}]{shampine2000solving}%
  \BibitemOpen
  \bibfield  {author} {\bibinfo {author} {\bibfnamefont {L.~F.}\ \bibnamefont
  {Shampine}}, \bibinfo {author} {\bibfnamefont {S.}~\bibnamefont {Thompson}},\
  and\ \bibinfo {author} {\bibfnamefont {J.}~\bibnamefont {Kierzenka}},\
  }\bibfield  {title} {\bibinfo {title} {Solving delay differential equations
  with dde23},\ }\href@noop {} {\bibfield  {journal} {\bibinfo  {journal}
  {http://www.runet.edu/~thompson/webddes/tutorial.pdf}\ } (\bibinfo {year}
  {2000})}\BibitemShut {NoStop}%
\bibitem [{\citenamefont {Subramanian}\ \emph {et~al.}(2010)\citenamefont
  {Subramanian}, \citenamefont {Mariappan}, \citenamefont {Sujith},\ and\
  \citenamefont {Wahi}}]{subramanian2010bifurcation}%
  \BibitemOpen
  \bibfield  {author} {\bibinfo {author} {\bibfnamefont {P.}~\bibnamefont
  {Subramanian}}, \bibinfo {author} {\bibfnamefont {S.}~\bibnamefont
  {Mariappan}}, \bibinfo {author} {\bibfnamefont {R.~I.}\ \bibnamefont
  {Sujith}},\ and\ \bibinfo {author} {\bibfnamefont {P.}~\bibnamefont {Wahi}},\
  }\bibfield  {title} {\bibinfo {title} {Bifurcation analysis of thermoacoustic
  instability in a horizontal {R}ijke tube},\ }\href@noop {} {\bibfield
  {journal} {\bibinfo  {journal} {International Journal of Spray and Combustion
  Dynamics}\ }\textbf {\bibinfo {volume} {2}},\ \bibinfo {pages} {325}
  (\bibinfo {year} {2010})}\BibitemShut {NoStop}%
\bibitem [{\citenamefont {Thomas}\ \emph {et~al.}(2018)\citenamefont {Thomas},
  \citenamefont {Mondal}, \citenamefont {Pawar},\ and\ \citenamefont
  {Sujith}}]{thomas2018effect}%
  \BibitemOpen
  \bibfield  {author} {\bibinfo {author} {\bibfnamefont {N.}~\bibnamefont
  {Thomas}}, \bibinfo {author} {\bibfnamefont {S.}~\bibnamefont {Mondal}},
  \bibinfo {author} {\bibfnamefont {S.~A.}\ \bibnamefont {Pawar}},\ and\
  \bibinfo {author} {\bibfnamefont {R.~I.}\ \bibnamefont {Sujith}},\ }\bibfield
   {title} {\bibinfo {title} {Effect of time-delay and dissipative coupling on
  amplitude death in coupled thermoacoustic oscillators},\ }\href@noop {}
  {\bibfield  {journal} {\bibinfo  {journal} {Chaos: An Interdisciplinary
  Journal of Nonlinear Science}\ }\textbf {\bibinfo {volume} {28}},\ \bibinfo
  {pages} {033119} (\bibinfo {year} {2018})}\BibitemShut {NoStop}%
\bibitem [{\citenamefont {Jegal}\ \emph {et~al.}(2019)\citenamefont {Jegal},
  \citenamefont {Moon}, \citenamefont {Gu}, \citenamefont {Li},\ and\
  \citenamefont {Kim}}]{jegal2019mutual}%
  \BibitemOpen
  \bibfield  {author} {\bibinfo {author} {\bibfnamefont {H.}~\bibnamefont
  {Jegal}}, \bibinfo {author} {\bibfnamefont {K.}~\bibnamefont {Moon}},
  \bibinfo {author} {\bibfnamefont {J.}~\bibnamefont {Gu}}, \bibinfo {author}
  {\bibfnamefont {L.~K.~B.}\ \bibnamefont {Li}},\ and\ \bibinfo {author}
  {\bibfnamefont {K.~T.}\ \bibnamefont {Kim}},\ }\bibfield  {title} {\bibinfo
  {title} {Mutual synchronization of two lean-premixed gas turbine combustors:
  {P}hase locking and amplitude death},\ }\href@noop {} {\bibfield  {journal}
  {\bibinfo  {journal} {Combustion and Flame}\ }\textbf {\bibinfo {volume}
  {206}},\ \bibinfo {pages} {424} (\bibinfo {year} {2019})}\BibitemShut
  {NoStop}%
\end{thebibliography}%

\end{document}